%% file: EbyEFluc_v1.tex
\documentclass[ALICE,manyauthors]{cernphprep}
\usepackage[comma,square,numbers,sort&compress]{natbib}
\usepackage{hyperref}
\usepackage{lineno}
\usepackage{xspace}
\usepackage{xcolor}
\usepackage{amssymb,amsmath}
\usepackage{MnSymbol}
\usepackage[T1]{fontenc}
\usepackage{orcidlink}

\begin{document}
\input{commands.tex}

\begin{titlepage}
\PHyear{2024}       
\PHnumber{297}      
\PHdate{12 November}  

\title{System size and energy dependence of the mean transverse momentum fluctuations at the LHC }

\ShortTitle{System size and energy dependence of the $\langle p_{\rm{T} }\rangle$ fluctuations at the LHC}   

\Collaboration{ALICE Collaboration\thanks{See Appendix~\ref{app:collab} for the list of collaboration members}}
\ShortAuthor{ALICE Collaboration} 

\begin{abstract}
Event-by-event fluctuations of the event-wise mean transverse momentum, \mpt, of charged particles produced in proton--proton (pp) collisions at \s = 5.02\,\TeV, Xe--Xe collisions at  \snn= 5.44\,\TeV, and Pb--Pb collisions at \snn= 5.02\,\TeV are studied using the ALICE detector based on the integral correlator $\llangle  \Delta p_{\rm T}\Delta p_{\rm T}\rrangle $. The correlator strength is found to decrease monotonically with increasing produced charged-particle multiplicity measured at midrapidity in all three systems.  In Xe--Xe and Pb--Pb collisions, the multiplicity dependence of the correlator deviates significantly from a simple power-law scaling as well as from the predictions of the HIJING and AMPT models. The observed deviation from power-law scaling is expected from transverse radial flow in semicentral to central Xe--Xe and Pb--Pb collisions.  In pp collisions, the correlation strength is also studied by classifying the events based on the transverse spherocity, $S_0$, of the particle production at midrapidity, used as a proxy for the presence of a pronounced back-to-back jet topology. Low-spherocity (jetty) events feature a larger correlation strength than those with high spherocity (isotropic). The strength and multiplicity dependence of jetty and isotropic events are well reproduced by calculations with the PYTHIA 8 and EPOS LHC models.
\end{abstract}
\end{titlepage}

\setcounter{page}{2} 


\section{Introduction} 
Studies of event-by-event fluctuations of event-wise observables measured in heavy-ion collisions are of great interest given that they probe the phase transition from quark--gluon plasma (QGP)  to hadron gas (HG)~\cite{Jeon:2003gk,Heiselberg:2000fk,Shuryak:1997yj,Stephanov:1998dy,Stephanov:1999zu,Dumitru:2000in,Fodor:2004nz,ALICE:2022wpn}. Of particular interest are fluctuations of the average transverse momentum ($p_{\rm T}$) of particles measured event-by-event in a specific kinematic range. 
These fluctuations are expected to be sensitive to energy fluctuations and, arguably, temperature variations of the matter produced in these collisions. In turn, the magnitude of these fluctuations is nominally proportional to the heat capacity of the hot medium, which is governed by the strong force as described by quantum chromodynamics (QCD)~\cite{Stodolsky:1995ds,Basu:2016ibk}. As such, the temperature fluctuations are predicted to sharply increase in the vicinity of the critical point and near a cross-over phase transition boundary,  as a result of a rapid change in the heat capacity of the medium near this  boundary~\cite{Stodolsky:1995ds}. Fluctuations of the mean $p_{\rm T}$, \mpt, are also highly sensitive to the presence of collective effects and the onset of thermalization in small systems.
Measurements of \mpt fluctuations are thus of great interest in the study of the hot and dense matter produced in heavy-ion collisions~\cite{Shuryak:1997yj,Stephanov:1998dy,Stephanov:1999zu}. 

A significant number of measurements of  \mpt fluctuations have already been carried out at SPS~\cite{NA49:1999inh,CERES:2003sap,NA49:2008fag,Nijs:2021clz} as well as at RHIC energies~\cite{STAR:2006rqs,STAR:2005xhg,STAR:2005vxr,STAR:2003cbv,STAR:2019dow,STAR:2024wgy,PHENIX:2003ccl,PHENIX:2002aqz} based on a variety of observables. Traditionally, the measurement of $\langle p_{\rm T} \rangle$ fluctuations is carried out in terms of the two-particle correlator, $\llangle \Delta p_{\rm T}\Delta p_{\rm T}\rrangle$, which measures the particle momentum correlations based on deviates of $p_{\rm T}$  relative to $\langle p_{\rm T} \rangle$ (discussed in Sec.~\ref{sec:methodolgy}). The ALICE Collaboration reported 
a measurement in Pb--Pb collisions at a center-of-mass energy per nucleon pair $\sqrt{s_{\rm NN}}$ = 2.76 TeV~\cite{ALICE:2014gvd} using the $\llangle \Delta p_{\rm T}\Delta p_{\rm T}\rrangle$ correlator. These prior measurements  identified the presence of finite dynamical fluctuations corresponding to non-vanishing $\llangle  \Delta p_{\rm T}\Delta p_{\rm T}\rrangle $ correlations and additionally studied the evolution of the strength of the correlator with collision centrality. The interpretation of the observed fluctuations in terms of temperature fluctuations is, however, challenged by various considerations which are discussed below.

In small collision systems, one expects that the magnitude of the correlator $\llangle  \Delta p_{\rm T}\Delta p_{\rm T}\rrangle $ should be primarily  determined by elementary particle production processes such as string fragmentation, hadronic resonance decays, and jets. 
The overall correlation strength measured in small collision systems should thus depend on the relative abundance of these processes and the relative strengths of the correlator for each of these processes. Correlations in large collision systems, on the other hand, should additionally depend on the number of individual nucleon--nucleon (or parton--parton) collisions and whether these produce collective phenomena or feature rescatterings of the particles they produce.  For collisions involving independent nucleon--nucleon collisions, one expects that the strength of the correlator should evolve in inverse proportion to the number of sources of correlated particles, which is generally expected in a dilution scenario. The dilution  results from  the superposition of approximately independent particle-emitting sources, i.e.,  independent  nucleon--nucleon collisions with no rescatterings of secondaries~\cite{Pruneau:2002yf,Sharma:2008qr}. 
This translates into an inverse dependence of the magnitude of the correlator on the average density of charged particles, $\langle {\rm d}N_{\rm{ch}}/{\rm d}\eta\rangle$, produced in a given interval of collision centrality and pseudorapidity ($\eta$). Although prior observations of $\llangle  \Delta p_{\rm T}\Delta p_{\rm T}\rrangle $ at  RHIC and LHC have shown that the magnitude of this correlator decreases monotonically from peripheral to central collisions, a sizable deviation from the $\langle {\rm d}N_{\rm{ch}}/{\rm d}\eta\rangle$ scaling behavior was observed in semicentral to central collisions of large systems~\cite{STAR:2005vxr,ALICE:2014gvd,ALICE:2023tej}. Fluctuations of the system energy (temperature) can evidently contribute to additional  transverse momentum fluctuations and to a relative increase  of $\llangle  \Delta p_{\rm T}\Delta p_{\rm T}\rrangle $,  but a number of  other mechanisms could potentially also explain the observed behavior. These include the onset of collectivity and thermalization~\cite{Voloshin:2003ud,Gavin:2003cb}, string percolation~\cite{Ferreiro:2003dw}, as well as initial-state energy density fluctuations~\cite{Gavin:2011gr,Gavin:2012if,Alver:2010gr}. 
Among these, the role of radial flow, well established from measurements of single-particle $p_{\rm T}$ distributions, may explain much of the deviation from inverse density scaling~\cite{Voloshin:2003ud}. 
However, whether this deviation can be understood quantitatively on the basis of the aforementioned scenarios remains an open question. It is thus of interest to
compare the collision centrality dependence of  $\llangle  \Delta p_{\rm T}\Delta p_{\rm T}\rrangle $ observed in large  collision systems with that found in elementary proton--proton (pp) collisions. As stated above, in nucleus--nucleus (A--A) collisions, one expects that the scaling is primarily driven by the number of (binary) nucleon--nucleon (or possibly parton--parton) interactions resulting in an approximate linear scaling with the total produced multiplicity density. Such scaling is seemingly not expected in pp collisions, but one can nonetheless expect that the magnitude of $\llangle \Delta p_{\rm T}\Delta p_{\rm T}\rrangle$ should evolve with the number of underlying correlated sources, whether these arise from string fragmentation,   jet production, or multipartonic interactions.
One must then examine in detail how the strength of  $\llangle \Delta p_{\rm T}\Delta p_{\rm T}\rrangle$ evolves with the particle density in both heavy-ion and pp collisions. Additionally, since stronger correlations are expected 
from the collimated particle production arising from the hadronization of jets, it is also of interest to compare the strength of the correlator in pp collisions by separating events with a pronounced back-to-back jet topology from events featuring approximately transversely isotropic particle distributions. This particular study is performed based on the transverse spherocity variable known to be sensitive to the transverse event shape~\cite{ALICE:2019dfi,Banfi:2010xy,Ortiz:2017jho},  which gives information on how the particles are distributed perpendicularly to the collision axis.

This paper presents measurements of event-by-event fluctuations of the event-wise mean transverse momentum, $\langle p_{\rm{T} }\rangle$, of charged particles produced in pp collisions at \s = 5.02 TeV, Xe--Xe collisions at \snn= 5.44\,\TeV, and Pb--Pb collisions at \snn= 5.02\,\TeV as a function of charged-particle multiplicity recorded using the ALICE detector at the LHC.  The primary goal of the measurements is to examine how the strength of the $\llangle \Delta p_{\rm T}\Delta p_{\rm T}\rrangle $ correlator evolves with the collision energy by comparing to previous results from Pb–Pb collisions at \s = 2.76 TeV and the collision system size, and determine whether this evolution can be understood quantitatively based on existing models. Additionally, since the presence
of jet constituents is likely to influence the magnitude of the measured correlations, particularly in small collision systems, this work also includes an analysis of the strength of the $\llangle \Delta p_{\rm T}\Delta p_{\rm T}\rrangle $  correlator in pp collisions based on the transverse shape of events (discussed in Sec.~\ref{sec:spherocity}).

The paper is organized as follows: Sec.~\ref{sec:methodolgy} presents a summary of the techniques used to evaluate the transverse momentum correlator  $\llangle  \Delta p_{\rm T}\Delta p_{\rm T}\rrangle $ and derived quantities used in this work. It also includes a short discussion of the definition and techniques used  towards measurements of the evolution of the  $\llangle  \Delta p_{\rm T}\Delta p_{\rm T}\rrangle $ correlation strength with the transverse event shape in pp collisions. Details of the experimental methods  and techniques used to determine systematic uncertainties are discussed in Sec.~\ref{sec:Experimental Methods}. The main results are presented in Sec.~\ref{sec:results} and compared with those of previous measurements and model predictions. The paper is concluded with a summary in Sec.~\ref{sec:summary}.
 
\section{Observable definitions }
\label{sec:methodolgy}

The formal definition of the $\llangle  \Delta p_{\rm T}\Delta p_{\rm T}\rrangle $ correlator in terms of two-particle density is introduced in Sec.~\ref{sec:methodolgy:1}. In this work, $\llangle  \Delta p_{\rm T}\Delta p_{\rm T}\rrangle $ is measured based on an event-wise estimator and the method of moments presented in Sec.~\ref{sec:methodolgy:2}. The transverse spherocity estimator is defined in Sec.~\ref{sec:spherocity}.

\subsection{Definition of  \texorpdfstring{$\llangle  \Delta p_{\rm T}\Delta p_{\rm T}\rrangle $}\xspace}
\label{sec:methodolgy:1}

Nominally, studies of average $p_{\rm T}$ fluctuations are carried out based on the integral correlator $\llangle  \Delta p_{{\rm T}1}\Delta p_{{\rm T}2}\rrangle$ (see Refs.~\cite{Voloshin:1999yf,Voloshin:2002ku,Voloshin:2003ud}) defined according to the following formula \begin{equation}
\label{eq:DptDptDef}
\llangle  \Delta p_{{\rm T}1}\Delta p_{{\rm T}2}\rrangle  \equiv  \frac{\int 
  \rho_2(p_{{\rm T}1},p_{{\rm T}2})\Delta p_{{\rm T}1}\Delta p_{{\rm T}2} \ {\rm d}p_{{\rm T}1}{\rm d}p_{{\rm T}2}}{\int \rho_2(p_{{\rm T}1},p_{{\rm T}2}){\rm d}p_{{\rm T}1}{\rm d}p_{{\rm T}2}},
\end{equation}
where $\rho_2(p_{{\rm T}1},p_{{\rm T}2})$ represents a two-particle density. This function is expressed in terms of the transverse momenta $p_{{\rm T}1}$ and $p_{{\rm T}2}$ of two particles.

The term  $\Delta p_{{\rm T}i}=p_{{\rm T}i}-\llangle p_{\rm T}\rrangle$, where $i=1,\, 2$, represents the transverse momentum deviates of particles 1 and 2, of a given pair, relative to  the inclusive average $\llangle   p_{\rm T}
\rrangle $.
The inclusive average $\llangle   p_{\rm T} \rrangle $ is defined according to 
\begin{equation}
\llangle   p_{\rm T}\rrangle  \equiv \frac{\int \rho_1(p_{\rm T})p_{\rm T}{\rm d}p_{\rm T}}{\int \rho_1(p_{\rm T}){\rm d}p_{\rm T}},
\end{equation}
in which $\rho_1(p_{\rm T})$ is the inclusive single-particle density.  

\subsection{Measurement method}
\label{sec:methodolgy:2}

The measurements of  $\llangle  \Delta p_{{\rm T}1}\Delta p_{{\rm T}2}\rrangle $  are obtained  based on an  event-wise statistical estimator~\cite{Voloshin:2002ku,ALICE:2014gvd,Bhatta:2021qfk} defined according to
\begin{eqnarray}
\label{eq:DptDptEstimator}
\llangle  \Delta p_{{\rm T}1}\Delta p_{{\rm T}2} \rrangle = \bigg\langle\frac{\sum_{i,j=1 ; i\ne j}^{N_{\rm{ch}}}(p_{{\rm T}i}-\llangle p_{\rm T} \rrangle)(p_{{\rm T}j}-\llangle p_{\rm T}\rrangle)}{ N_{\rm{ch}}(N_{\rm{ch}}-1)}\bigg\rangle,
\end{eqnarray}
with the event-wise average transverse momentum 
\begin{eqnarray}
\llangle p_{\rm T} \rrangle= 
\frac{1}{N_{\rm{ch}}}
 \sum
 \limits_{i=1}^{N_{\rm ch}}p_{{\rm T}i}. 
\end{eqnarray}
In these equations, $N_{\rm{ch}}$ represents the total number of charged particles measured within a single event, and $p_{{\rm {T}},i}$ and $p_{{\rm T},{j} }$ denote the transverse momenta of the $i^{\rm th}$ and $j^{\rm th}$ particles, respectively, with $i,j=1,\ldots, N_{\rm{ch}}$, and $i \neq j$ to avoid self-correlations. The average is said to be an  event-wise average because the sum of the product of deviates is divided by the number of pairs of particles in each event. The angle bracket, $\llangle  O \rrangle $, represents the average of the event-wise  observable $\langle O\rangle$ computed over an event ensemble of interest. 
 In this analysis, values of the correlator $\llangle  \Delta p_{{\rm T}1}\Delta p_{{\rm T}2} \rrangle $  were determined for minimum bias event samples and for specific classes of the events selected based on their charged-particle multiplicity measured in forward and backward detectors (see Sec.~\ref{sec:Experimental Methods}). Additionally, the pp collisions were categorized into event subsets based on a measurement of their transverse spherocity defined in Sec.~\ref{sec:spherocity}.

Computationally, it is advantageous to reformulate the analysis of $\llangle  \Delta p_{{\rm T}1}\Delta p_{{\rm T}2} \rrangle $  with the introduction of an event-wise variable $Q_n$ defined according to 
\begin{eqnarray}
Q_{ n}=\sum_{i=1}^{N_{\rm{ch}}} (p_{{\rm T}i})^n,
\end{eqnarray}
where $p_{{\rm T}i}$ represents the transverse momentum of particles, $i=1, \ldots, N_{\rm{ch}}$,   used in the measurement of $\llangle  \Delta p_{{\rm T}1}\Delta p_{{\rm T}2} \rrangle $.
One verifies that $\llangle  \Delta p_{{\rm T}1}\Delta p_{{\rm T}2} \rrangle $ can be readily computed, according to~\cite{Giacalone:2020lbm}, as
\begin{eqnarray}
 \llangle  \Delta p_{{\rm T}1}\Delta p_{{\rm T}2} \rrangle = \bigg\langle\frac{(Q_1)^2-Q_2}{N_{\rm{ch}}(N_{\rm{ch}}-1)}\bigg\rangle - \bigg\langle\frac{Q_1}{N_{\rm{ch}}}\bigg\rangle^2.
\end{eqnarray}

This analytic approach~\cite{Giacalone:2020lbm} simplifies the computations, thus significantly reducing the analysis time, especially in high-multiplicity events. In order to study the particle-density dependence of the correlator and minimize smearing effects associated with broad bin widths, the analysis is performed in narrow intervals of the charged-particle multiplicity detected at forward rapidity (i.e., the  forward detector acceptance as described in Sec.~\ref{sec:Experimental Methods}).

In the absence of particle correlations, i.e., for purely Poissonian fluctuations of the event-wise \mpt, the correlator  $\llangle  \Delta p_{{\rm T}1}\Delta p_{{\rm T}2} \rrangle $ vanishes. However, it acquires a  finite value, either positive or negative, when the  transverse momenta of the produced particles are correlated. Note that both the numerator and the denominator of $\llangle  \Delta p_{{\rm T}1}\Delta p_{{\rm T}2} \rrangle $ are proportional to the square of the particle detection efficiency making $\llangle  \Delta p_{{\rm T}1}\Delta p_{{\rm T}2} \rrangle $ robust against particle losses~\cite{ALICE:2014gvd}, i.e., efficiencies approximately cancel out in measurements of $\llangle  \Delta p_{{\rm T}1}\Delta p_{{\rm T}2} \rrangle $. However, the cancellation is not perfect, particularly if the detection efficiency depends on the $p_{\rm T}$ of particles.

The $\llangle  \Delta p_{{\rm T}1}\Delta p_{{\rm T}2} \rrangle $ correlator measures particle momentum correlations based on deviates relative to the mean momentum. In heavy-ion collisions, the mean momentum is known to depend in part on collective effects and more particularly, radial flow~\cite{ALICE:2013rdo}. One thus expects $\llangle  \Delta p_{{\rm T}1}\Delta p_{{\rm T}2} \rrangle $ to depend on the magnitude of radial flow. This dependence can be largely suppressed by formulating the results in terms of the dimensionless quantity  $\sqrt{ \llangle  \Delta p_{{\rm T}1}\Delta p_{{\rm T}2} \rrangle }/{\llangle p_{\rm T} \rrangle}$, indicated as the normalized transverse momentum correlator.
The use of a dimensionless observable features a number of additional advantages: independence from uncertainties on the momentum scale, partial independence from the magnitude of  $\langle p_{\rm T}\rangle$, and further reduction of sensitivity to the dependence of the particle detection and reconstruction efficiency on the transverse momentum~\cite{ALICE:2014gvd}. 
The results reported here are thus presented in terms of $\sqrt{ \llangle  \Delta p_{{\rm T}1}\Delta p_{{\rm T}2} \rrangle }/{\llangle p_{\rm T} \rrangle}$, in lieu of $\llangle  \Delta p_{{\rm T}1}\Delta p_{{\rm T}2} \rrangle $. Note that at variance with prior notations~\cite{STAR:2005vxr}, the double bracket notation is used to clearly denote that the correlator is measured as an event ensemble average of  the average  pair-wise  $\Delta p_{\rm T}\Delta p_{\rm T}$ measured event-by-event.  The measurement carried out in this work is otherwise equivalent to those reported earlier in Ref.~\cite{ALICE:2014gvd}. 

\subsection{Definition of the transverse spherocity}
\label{sec:spherocity}

In small collision systems, particularly in pp collisions, the many specific processes contributing to particle production have varying (fluctuating) contributions from one collision to another. Some collisions may thus feature sizable contributions from minijets (created by hard QCD scatterings at intermediate \pt) or jets, while other may be dominated by ``soft" multipartonic interactions. Given that, on average, these processes feature different momentum scales and produced multiplicities, it is of interest to examine the relative role they play in influencing the strength of the $\llangle \Delta p_{\rm T1}\Delta p_{\rm T2} \rrangle$ correlator. This is accomplished by further classifying pp collisions based on their transverse shape estimated with the transverse spherocity observable, $S_{0}$~\cite{ALICE:2023bga}. 

The observable, $S_{0}$, is defined according to 
\begin{align}
S_{0}=\frac{\pi^2}{4}
\min_{\hat{n}=(n_x,n_y,0)}
\left(
\frac
{  \sum\limits_{i=1}^{N_{\rm ch}} \hat p_{{\rm T}_i}\ \times\ \hat n}
{  \sum\limits_{i=1}^{N_{\rm ch}}  \hat p_{{\rm T}_i} }
\right)^2,
\end{align}
where $\hat p_{{\rm T}_i}$ represents transverse momentum unit vectors.
The orientation of  $\hat n$, a two-dimensional unit vector of momentum in the transverse ($xy$) plane, perpendicular to the beam axis, is chosen such that $S_{0}$ is minimized on an event-by-event basis. By construction, $S_{0}$ ranges from 0 for pencil-like (jetty) events to a maximum of 1 for circularly symmetric events in the transverse plane,  i.e., transversely isotropic events. In this analysis, the $p_{\rm T}$-unweighted definition of transverse spherocity ($p_{\rm T} = 1$), already employed in prior ALICE works~\cite{ALICE:2023bga}, is used to quantify the event topology in the transverse plane because it reduces biases otherwise introduced with the original spherocity definition~\cite{Banfi:2010xy}.

\section{Data analysis}
\label{sec:Experimental Methods}

 The data samples used in this analysis were collected by the ALICE experiment during the data-taking periods with pp collisions at \s= 5.02 and 13 TeV in 2015, Pb--Pb collisions at \snn = 5.02 TeV in 2015, and Xe--Xe collisions at  \snn= 5.44 TeV in 2017.
The data were acquired with a minimum bias trigger requiring coincident signals in the two scintillator arrays of the \VZERO detector covering forward (\VZEROA, $2.8<\eta <5.1$) and backward (\VZEROC,$-3.7< \eta< -1.7$) pseudorapidity intervals. The \VZERO detector helps to reject the beam-induced background via \VZERO timing cuts.
Detailed descriptions of the ALICE detectors, its components, and their performance, have been reported in Refs.~\cite{ALICE:2008ngc,Abelev:2014ffa}.

Charged-particle multiplicities measured with the \VZERO detectors were additionally used to divide the measured datasets into several multiplicity classes expressed as percentiles of the total hadronic cross section:  eleven in Pb--Pb~\cite{ALICE:2015juo} and Xe--Xe~\cite{ALICE:2018cpu}  collisions, and nine in pp collisions. These classes were used to characterize the $\langle p_{\rm T} \rangle$ fluctuations with respect to the produced charged-particle multiplicity in pp collisions. In heavy-ion collisions, the charged-particle multiplicity is related to the impact parameter, which is the distance in the transverse plane between the centers of the colliding nuclei, of Xe--Xe and Pb--Pb collisions. 

Measurements of $\langle p_{\rm T} \rangle$ fluctuations were based on charged-particle tracks reconstructed with the ITS and the Time Projection Chamber (TPC). The analysis was further restricted to particles emitted within the pseudorapidity range $|\eta| < 0.8$  and the transverse momentum range $0.15 < p_{\rm T}< 2$ GeV/$c$.  The selection on $p_{\rm T}$ is designed to focus the analysis of soft particle production in the ``bulk" while minimizing contributions from the fragmentation of jets. The pseudorapidity range is selected to ensure uniform particle detection efficiency.

The events with a reconstructed primary vertex within  10 cm of the nominal interaction point along the beam direction ($|V_z| <10$  cm) are chosen to ensure uniform acceptance in pseudorapidity in $|\eta| < 0.8$  for the ITS.
Additionally, events were considered in the analysis if at least one accepted charged particle contributed to the reconstruction of the primary vertex. Furthermore, events featuring more than one reconstructed primary interaction vertex were rejected to suppress the possibility of event pile-up. In all,  13 million, 1.4 million, and 104 million events passed the above criteria and were retained towards the analysis of \mpt fluctuations from \PbPb, \XeXe, and \pp ($\sqrt s$ = 5.02 TeV) collisions, respectively. 

Individual charged-particle tracks were also subjected to track-quality selection criteria and to specific selections to limit the analysis to primary particles, i.e., particles with a mean proper lifetime $\tau$
 larger than 1 cm/$c$, which are either produced directly in the interaction, or from decays of particles with $\tau$ smaller than 1 cm/$c$, restricted to decay chains leading to the interaction~\cite{ALICE-PUBLIC-2017-005}. In pp collisions, the analysis was limited to charged tracks with a minimum of  $N_{\rm{TPC}} =70$ reconstructed space points in the TPC, out of a maximum of 159,  whereas in
\PbPb and \XeXe collisions, a \pt-dependent cut, $N_{\rm{TPC}}$ = 70 + 1.5 $\times \frac{\pt}{\rm{GeV/\it c}}$,  was applied to further limit the probability of split tracks which is more relevant for events with high hit occupancy in the TPC. This occurs when a single particle is reconstructed as multiple separate tracks.
Additionally, in order to suppress secondary charged particles, the track distance-of-closest-approach (DCA) to the reconstructed primary interaction vertex was limited to $|d_{z}|<1$\,cm in the longitudinal direction and  $|d_{{xy}}|< 1$\,cm  in the transverse direction in \PbPb and \XeXe collisions. Finally, the DCA selection criteria,  $|d_{z}|<2$\,cm and $|d_{{xy}}|<0.0182$ + $0.0350$\,/${p_{\rm T}^{1.01}}$ cm, with $p_{\rm T}$ expressed in units of GeV/$c$ were used to minimize contamination from secondary particles in pp collisions~\cite{ALICE:2018vuu}.

Charged particles selected for the determination of $S_{0}$ were measured within the TPC and the ITS  and required to have transverse momentum $ p_{\rm T} > 0.15$ GeV/$c$  and lie within the pseudorapidity interval $|\eta| < 0.8$. The event shape-dependent analysis was further restricted to events featuring a minimum of five charged particles, $N_{\rm{ch}}\ge 5$, to ensure that the notion of transverse topology (a.k.a. transverse event shape) is meaningful. The analysis of $\llangle  \Delta p_{{\rm T}1}\Delta p_{{\rm T}2}\rrangle $ was carried out as a function of the strength of $S_0$ and is reported for  jetty events and isotropic events, based on two spherocity event classes corresponding to the lowest and top 20 percentile of all accepted events, respectively.

A comprehensive analysis was conducted to understand the impact of the detector response and analysis procedure on the measured observables, utilizing simulations based on different Monte Carlo event generators and on the  GEANT3~\cite{Brun:1082634} transport code,  including a detailed description of the ALICE detector components and their performance. Observables of interest were computed at the generator level, i.e., directly based on the output of event generators, and at the detector level, i.e., based on the output of the reconstruction of simulated events. Ratios of the detector-level and generator-level results were considered in what is known as a closure test. The generator level computation involves no particle losses and no resolution smearing,  whereas the reconstructed level includes both losses and smearing effects, as well as potential sources of contamination of the signal (poorly reconstructed tracks, tracks resulting from secondary particles, etc.).  The event generators used for the MC closure tests were HIJING~\cite{Wang:1991hta} for Pb--Pb and Xe--Xe collisions, and PYTHIA8~\cite{Skands:2014pea} for pp collisions. From the analysis of the closure test, it was found that the detector-level results matched those obtained at the generator level within 2\%. This difference was assigned as a systematic uncertainty, as discussed in Sec.~\ref{sec:Systematic uncertainties}.

The overall good closure obtained from the simulated data indicates, in particular, that the  observable
$\sqrt{ \llangle  \Delta p_{{\rm T}1}\Delta p_{{\rm T}2}\rrangle }$$/{\llangle  p_{\rm T} \rrangle }$ is robust against particle losses and the analysis reported thus does not include corrections for such losses.

In addition to HIJING and PYTHIA 8, the  AMPT~\cite{Lin:2004en} and EPOS LHC~\cite{Werner:2013tya} models  were used to compute the magnitude of the $\sqrt{ \llangle  \Delta p_{{\rm T}1}\Delta p_{{\rm T}2}\rrangle }$$/{\llangle  p_{\rm T} \rrangle }$ correlator and its evolution with collision centrality in Xe--Xe and Pb--Pb collisions to provide  insight into the interpretation of the data. The subsequent paragraphs provide a succinct overview of the models utilized for the different collision systems.

AMPT is a heavy-ion collision model that features partonic scattering and string fragmentation components in addition to a transport model.  It has had considerable successes in reproducing observables measured in heavy-ion collisions at both RHIC and LHC energies, such as the strength of anisotropic flow harmonics~\cite{Lin:2004en,Solanki:2012ne}.  However, it  has encountered mitigated success in the prediction of correlation and fluctuation observables~\cite{STAR:2010mib}. The data presented in this paper will enable further testing of the underlying physics hypotheses of the model. AMPT simulations were performed with default and string melting settings. In the default mode, the model assumes that hadrons are produced directly from strings via Lund string fragmentation whereas in the string melting mode, the model melts these strings into their constituent partons. Thus, all produced hadrons are decomposed into partons immediately after their formation. After the partonic phase, partons recombine to form hadrons through quark coalescence.

HIJING is a perturbative QCD inspired MC event generator for the study of jet and multiparticle production in high-energy pp and heavy-ion collisions.  The model includes multi-minijet production, nuclear shadowing of parton distribution functions, and mechanisms of jet interactions in a dense medium. In this analysis,  HIJING simulations are used for comparisons with results from heavy-ion collisions.

 PYTHIA is a MC event generator designed to simulate high-energy collisions between electrons, protons, photons, and heavy-nuclei. It features hard and soft interactions, sampling of parton distributions, initial- and final-state parton showers, multiparton interactions, as well as fragmentation and decays. Two tunes of PYTHIA are used in this analysis. PYTHIA 6 Perugia 2011 includes the revised set of parameters of flavor and fragmentation, which improves the overall description of  Tevatron data and the reliability of their extrapolations to LHC energies~\cite{Skands:2010ak}. The minimum bias and underlying event data from the LHC are taken into account in Perugia 2011 tune. PYTHIA 8 Monash tune includes a default parameterization of the model based on multiparton interactions and color reconnection mechanism. Following the hard scattering and parton showers, colored strings are formed with the final-state partons. PYTHIA 8 has a hadronization mechanism based on the fragmentation model which is followed by particle decays, which leads to the production of jets and the underlying event~\cite{Skands:2014pea}. 

The EPOS LHC model features particle production from core and corona components. Particle production in the corona is described in the context of the Parton-Based Gribov--Regge theory, whereas the core, expected to feature high parton densities, is described with ideal  hydrodynamics.  EPOS LHC is tuned to the LHC data via the color exchange mechanism of string excitation~\cite{Pierog:2013ria}.

\subsection{Systematic uncertainties}
\label{sec:Systematic uncertainties}

Several potential  sources of systematic uncertainties were considered including effects due to collision pile-up, contributions from tracks reconstructed with limited precision, secondary tracks, non-uniformity of the acceptance, and possible dependencies of the track reconstruction efficiency on the position of the primary vertex. Testing for these contributions was accomplished by studying the magnitude of the $\sqrt{ \llangle  \Delta p_{\rm T1}\Delta p_{\rm T2} \rrangle }/{\llangle p_{\rm T} \rrangle }$ correlator in response to variations of event and track quality selection criteria used in the analysis and also by comparing different data subsets. The systematic uncertainty for each source was determined based on the maximum difference between the results obtained with the default analysis configuration and with the different variations of the selection criteria.  

At the event level, the selections on the reconstructed position of the primary vertex along the beam axis ($V_z$), and on the presence of multiple reconstructed primary vertices to suppress events with collision pile-up were varied. Events with out-of-bunch pile-ups are identified and rejected exploiting the fact that the event multiplicity estimated from the signals measured with detectors with different readout times are affected differently by out-of-bunch pileup. Variables sensitive to multiplicity, extracted from detectors with different readout times, are used as tools to tag and remove events with out-of-bunch pile-ups. This  effectively removes most of the event pileup accumulated during the TPC drift and readout time. For the default configuration, a pile-up selection based on ITS and TPC clusters are applied. For systematic uncertainty, the selection is removed and the difference is conservatively quoted as systematic uncertainty due to pile-up. The $V_z$ selection is varied to $|V_z|<$ 8 cm and contributes to relative uncertainties smaller than 2.1\% for pp collisions, 3.7\% for Xe--Xe collisions, and 2.2\% in Pb--Pb collisions. The relative uncertainties due to remaining pile-up contamination are less than 2\% and 3.3\% for pp and Xe--Xe collisions, respectively, and vary from a minimum of 0.1\% to a maximum of 3.5\% for Pb--Pb collisions. At the track level, the track-quality and primary-particle selection criteria were varied including the longitudinal and transverse DCA of each track and the minimum number of TPC clusters required on a track. The maximum contribution to the systematic uncertainty due to the track selection criteria arises from the $d_{z}$ selection, and it is less than 5.5\%, 3.1\%, and 0.8\% in pp, Xe--Xe, and Pb--Pb collisions, respectively. Other relevant contributions to the systematic uncertainty on the track selection are due to the $d_{xy}$ cut (up to 4\% in pp collisions), the $d_{z}$ cut (up to 5.5\% in pp collisions), and the request of a minimum number of TPC clusters (up to 2.2\% in Xe--Xe collisions). The analysis of the closure test results showed that deviations between detector and generator levels do not exceed 2\% in all studied collision systems. The minor deviations from perfect closure are conservatively added to the systematic uncertainties. 
 The estimated values of the relative systematic uncertainties for the three collision systems are summarized in Table~\ref{Table1}. For the sources of uncertainty that depend on multiplicity a range of values is reported.
Individual contributions are summed in quadrature to obtain the  total systematic uncertainties.
We report the minimum and maximum values of systematic uncertainty due to different sources across all $\langle{\rm d}N_{\rm ch}/{\rm d}\eta \rangle$ values. The total systematic uncertainty gives a lower and upper limit across bins of $\langle{\rm d}N_{\rm ch}/{\rm d}\eta \rangle$.
The total systematic uncertainties are smaller than 7.1\%, 6.7\%, and 3.8\% for \pp, \XeXe and \PbPb collisions, respectively.

\begin{table} [!hpt]
\centering
\caption{Contributions to the relative (\%) systematic uncertainty on $\sqrt{ \llangle \Delta p_{{\rm T}1}\Delta p_{{\rm T}2} \rrangle}/{\llangle  p_{\rm T} \rrangle }$ of primary charged particles in \pp and \PbPb collisions at \snn $=$ 5.02\,\TeV and \XeXe collisions at \snn $=$ 5.44\, TeV. \\ }                
                \label{tab:sys}
                \begin{tabular}{|l |l |l |l |}
                \hline
                \textbf{Source of uncertainty}  & \pp, 5.02 \TeV & \XeXe, 5.44 \TeV & \PbPb, 5.02 \TeV\\ 
                \hline
                Vertex selection            & $<$ 2.1\%  & $<$ 3.7\% & $<$ 2.2\% \\
                Pile-up             &$<$ 2.0\%  &$<$ 3.3\% & 0.1--3.5\% \\
                No. of TPC clusters            &0.1--2.0\% & $<$ 2.2\%     & $<$ 0.5\% \\
                $d_{xy}$            &0.2--4.0\% & $<$ 1.5\%     & $<$ 0.3\% \\
                $ d_{z}$            &3.1--5.5\% & $<$ 3.1\%     & $<$ 0.8\% \\
                 MC closure             &1.3\% &1.8\%  &1.5\% \\
                \hline
                Total               &4.0--7.1\% & 1.8--6.7\%  &2.2--3.8\% \\
                \hline
                \end{tabular}

               \label{Table1}
\end{table}

\section{Results}
\label{sec:results}

Results of the study of the dependence of the magnitude of the  $\sqrt{ \llangle  \Delta p_{{\rm T}1}\Delta p_{{\rm T}2} \rrangle }/{\llangle  p_{\rm T} \rrangle }$ correlator on the produced particle density measured in pp, \XeXe, and \PbPb\ collision systems 
are presented in Sec.~\ref{sec:resultsVsMult}. The comparison of the measurements to theoretical predictions, and the energy dependence of the correlator are discussed in Secs.~\ref{sec:model} and ~\ref{sec:energy}, respectively. The dependence of  the  $\sqrt{ \llangle  \Delta p_{{\rm T}1}\Delta p_{{\rm T}2} \rrangle }/{\llangle  p_{\rm T} \rrangle }$ on the event  spherocity $S_0$ measured in pp collisions is discussed in Sec.~\ref{sec:resultsVsSo}

\subsection{System size dependence of event-by-event 
\texorpdfstring{$\langle p_{\rm{T} }\rangle$}\xspace~ 
fluctuations}
\label{sec:resultsVsMult}
\begin{figure}[ht!]
    \begin{center}
\includegraphics[width = 0.5\textwidth]{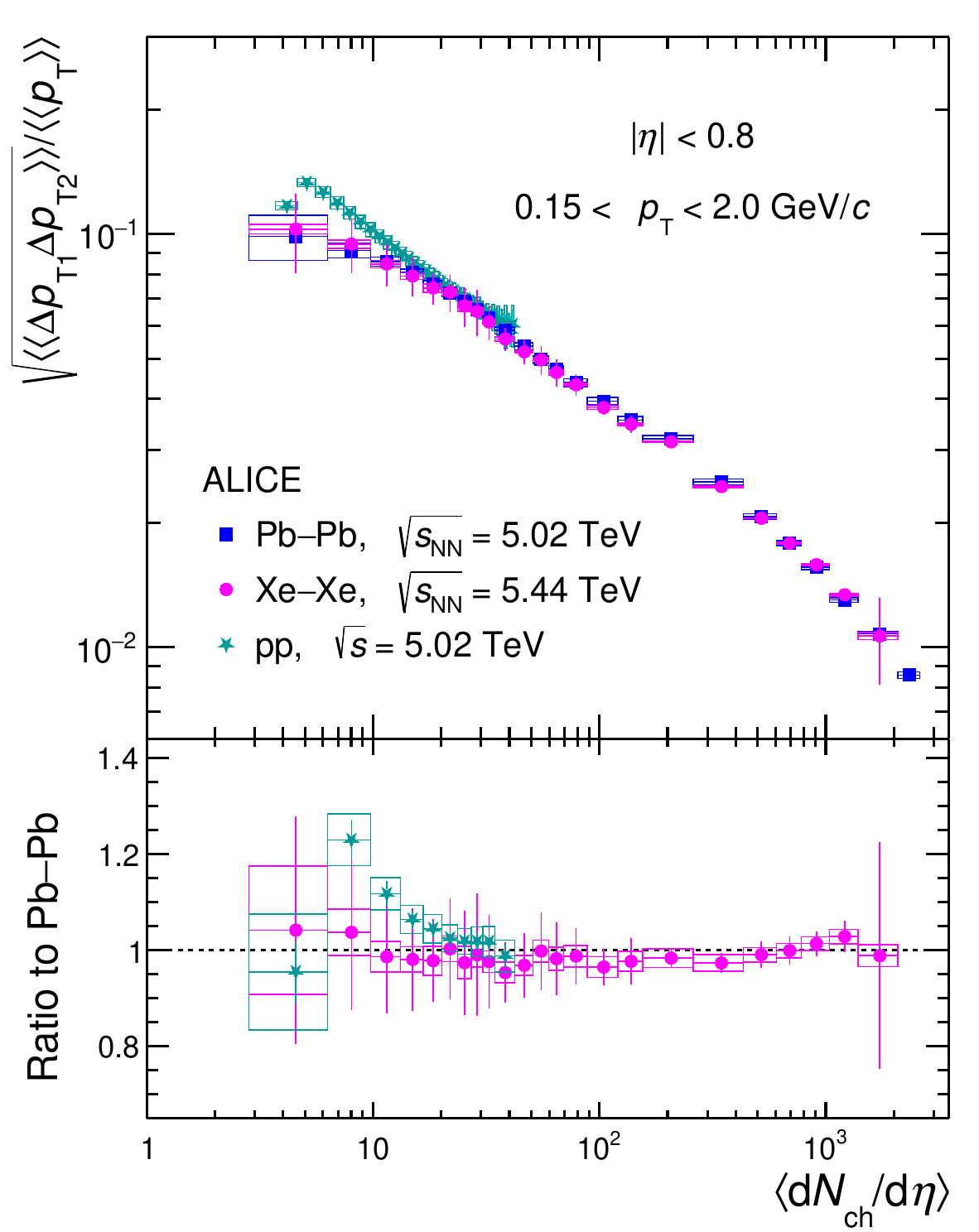} 

\end{center}
\caption{(top) Normalized transverse momentum correlator, $\sqrt{ \llangle  \Delta p_{{\rm T}1}\Delta p_{{\rm T}2} \rrangle }$$/\llangle  p_{\rm T} \rrangle $,  
    shown as a function of the charged-particle multiplicity density, $\langle{\rm d}N_{\rm ch}/{\rm d}\eta \rangle$, measured in the pseudorapidity range $|\eta| < 0.8$
    in \pp collisions at \s = 5.02 TeV, \XeXe collisions at \snn = 5.44 TeV, and \PbPb collisions at \snn = 5.02 TeV; (bottom) Ratio of values of $\sqrt{ \llangle  \Delta p_{{\rm T}1}\Delta p_{{\rm T}2} \rrangle }$$/\llangle  p_{\rm T} \rrangle $ measured in \pp and \XeXe collisions to those observed in \PbPb collisions. Statistical and systematic uncertainties are represented by vertical bars and boxes, respectively.}
    \label{fig:1}
\end{figure}

The event-by-event~$\langle p_{\rm{T} }\rangle$ fluctuations are reported based on the  
two-particle correlator  $\sqrt{ \llangle  \Delta p_{{\rm T}1}\Delta p_{{\rm T}2} \rrangle }$$/\llangle  p_{\rm T} \rrangle $ defined in Eq.~(\ref{eq:DptDptDef}) and computed according to Eq.~(\ref{eq:DptDptEstimator}). The top panel of Fig.~\ref{fig:1} presents the magnitude of $\sqrt{ \llangle  \Delta p_{\rm T1}\Delta p_{\rm T2} \rrangle }/{\llangle  p_{\rm T} \rrangle }$  measured as a function of the pseudorapidity density of charged particles produced in the collision, $\langle {\rm d}N_{\rm{ch}}/{\rm d}\eta\rangle$, determined in the kinematic range  $0.15< p_{\rm T}< 2$ GeV/$c$ and $|\eta|<0.8$, in \pp collisions at \s = 5.02 TeV, \XeXe collisions at \snn = 5.44 TeV, and \PbPb collisions at \snn = 5.02 TeV. The strength of  $\sqrt{ \llangle  \Delta p_{\rm T1}\Delta p_{\rm T2} \rrangle }/{\llangle  p_{\rm T} \rrangle }$ is evidently non-vanishing and exhibits an approximate power-law dependence on the produced charged-particle density. Fluctuations of the event-wise average momentum $\langle p_{\rm{T} }\rangle$ are accordingly non-Poissonian and exhibit a strong dependence on the particle density in all three collision systems studied. These new results confirm and corroborate prior observations of non-Poissonian fluctuations in heavy-ion collisions and make it possible to carry out a detailed study of the system size and energy dependence of the fluctuations~\cite{NA49:1999inh,CERES:2003sap,STAR:2003cbv,STAR:2005vxr,NA49:2008fag,ALICE:2014gvd,ALICE:2019smr,ATLAS:2019pvn,ATLAS:2024jvf}.

The correlator strength is observed to decrease by more than one order of magnitude with increasing multiplicity for Xe--Xe and Pb--Pb  collisions. One notes, however, that this multiplicity dependence cannot be described by a single power-law across the whole range of $\langle {\rm d}N_{\rm{ch}}/{\rm d}\eta\rangle$. One finds, indeed, that in both Pb--Pb (blue square) and Xe--Xe (magenta circle) collisions, the dependence can be characterized by three power-law regimes with  distinct slopes  in the ranges $3< \langle {\rm d}N_{\rm{ch}}/{\rm d}\eta\rangle<20$, $\ 20< \langle {\rm d}N_{\rm{ch}}/{\rm d}\eta\rangle <300$, and $ \langle {\rm d}N_{\rm{ch}}/{\rm d}\eta\rangle > 300$, respectively. This suggests that the strength of the correlation is influenced by several distinct  mechanisms (or system properties) from the most peripheral to the most central collisions considered in this study, as discussed in more detail in the following.  
 
The lower panel of Fig.~\ref{fig:1} displays the evolution of the ratio of the correlation strength measured in \pp and \XeXe collisions relative to that observed in  \PbPb collisions at $\sqrt{s_{\rm NN}}=5.02$ TeV. To compute the ratio, the pp and Xe--Xe results were rebinned to the $\langle {\rm d}N_{\rm{ch}}/{\rm d}\eta\rangle$ intervals used for Pb--Pb results, using a fit function. It is apparent that the magnitude of the correlators measured in \PbPb and \XeXe are consistent between each other and feature essentially the same dependence on $\langle {\rm d}N_{\rm{ch}}/{\rm d}\eta\rangle$. By contrast, however, the evolution of the correlator strength measured in pp collisions differs from that observed in the larger systems. 
Nevertheless, in pp collisions at the higher values ($>$20) of particle multiplicities, $\langle {\rm d}N_{\rm{ch}}/{\rm d}\eta\rangle$, the correlator strength is in very good agreement, within statistical uncertainties, to that reported in Pb–Pb and Xe–Xe at the same particle density.
For decreasing $\langle {\rm d}N_{\rm{ch}}/{\rm d}\eta\rangle$ the correlation strength measured in pp progressively deviates from the values observed for the larger systems. The scaling seen for Pb--Pb and Xe--Xe at low multiplicity, seems to be caused by the dilution of the correlator with increasing number of correlator sources which grow with participant number~\cite{Pruneau:2003ix}. At low multiplicities, large non-Gaussian fluctuations deviate from the expected linear scaling, unlike in pp collisions where the number of sources fluctuates less. In pp collisions, the scaling would be similar but in terms of the number of partonic collisions resulting in variations in MPI, mini jets, or jets that may lead to a smoothly varying number of correlated sources. One consequently observes an approximately $A^{-0.5}$ scaling.  The nuclear structures of Xe and Pb are distinct due to their different sizes, neutron-to-proton ratios, and shell structure. The Pb nucleus is notably heavier, and features a larger neutron skin, but is otherwise more spherical. Correlations between deformation, anisotropic flow, and transverse momentum have been reported by ATLAS~\cite{ATLAS:2022dov}. However, the current measurement does not enable us to readily discern effects associated with deviations from a spherical structure. Overall, one finds that the evolution of $\sqrt{ \llangle  \Delta p_{\rm T1}\Delta p_{\rm T2} \rrangle }/{\llangle p_{\rm T} \rrangle }$ with $\langle {\rm d}N_{\rm{ch}}/{\rm d}\eta\rangle$ follows similar trends in \PbPb and \XeXe whereas pp interactions reveals discrepancies at low multiplicities.

\subsection{Comparison to theoretical predictions}
\label{sec:model}
\begin{figure}[ht!]
    \begin{center}
    \includegraphics[width = 0.45\textwidth]
    {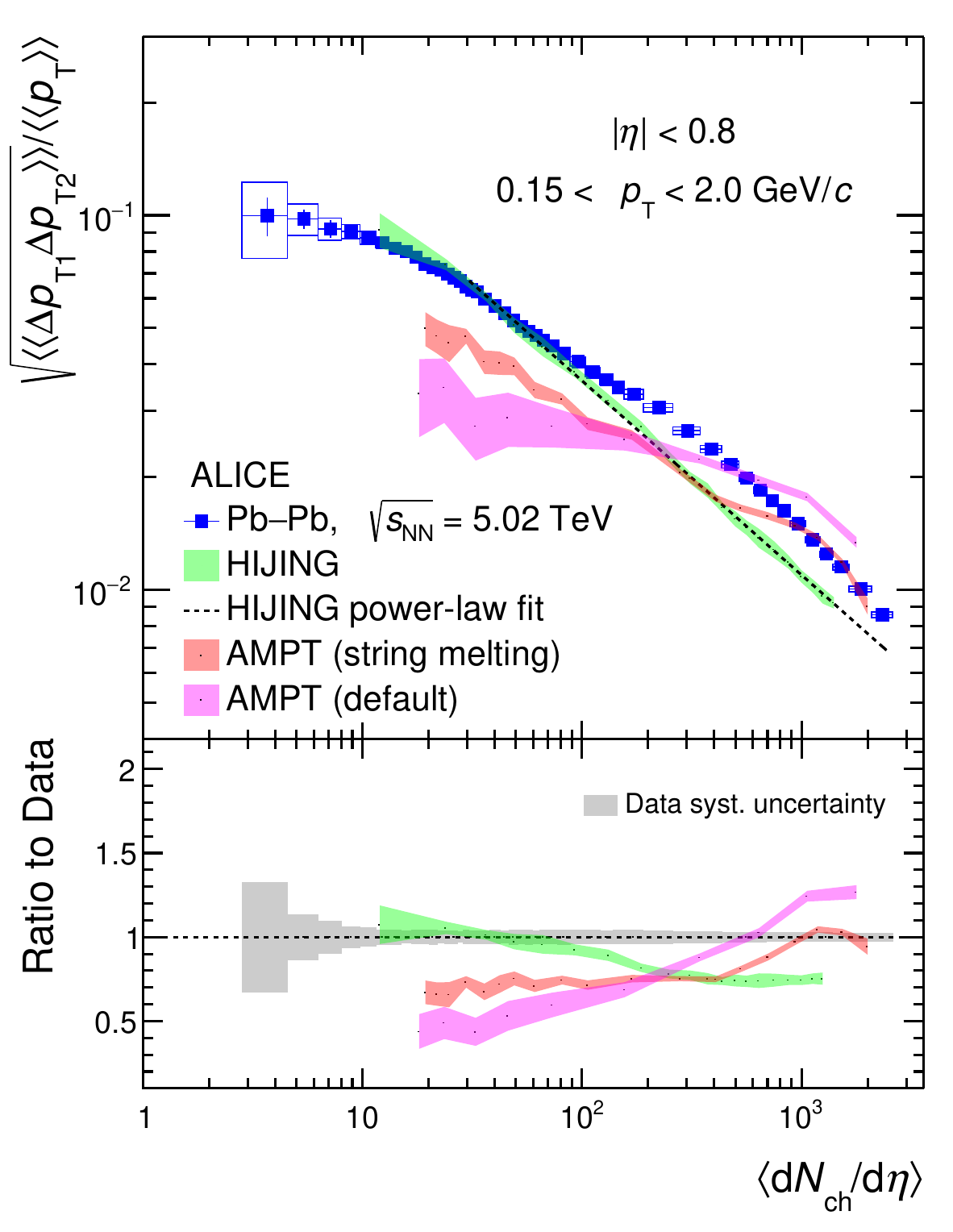}
      \includegraphics[width = 0.45\textwidth]
           {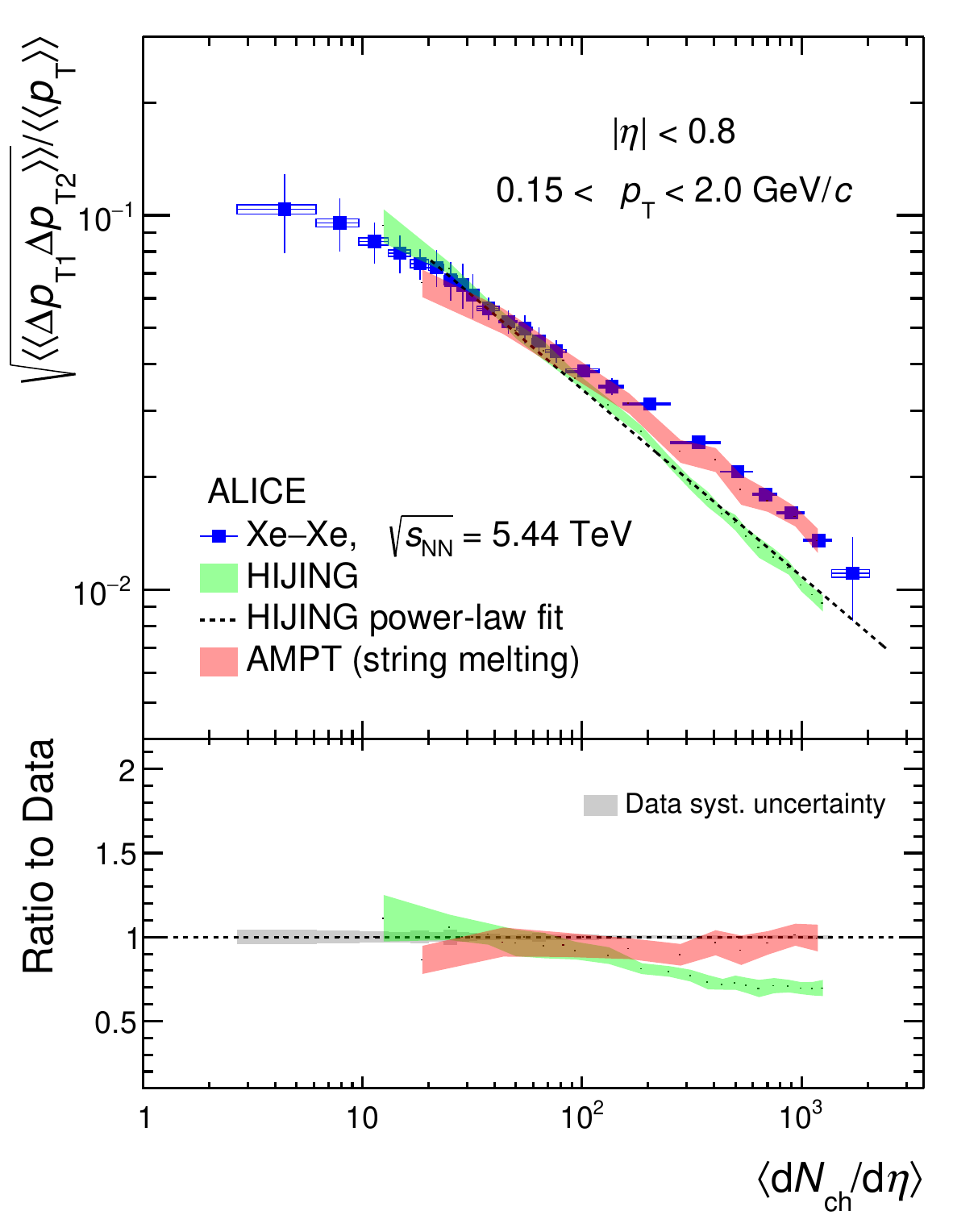}
    \end{center}
    \caption{Upper panels: Comparisons of the evolution of the  strength of  $\sqrt{ \llangle  \Delta p_{\rm T1}\Delta p_{\rm T2} \rrangle }/{\llangle p_{\rm T} \rrangle }$  with produced charged-particle multiplicity densities, $\langle {\rm d}N_{\rm ch}/{\rm d}\eta\rangle$,
    in \PbPb collisions at \snn = 5.02 TeV (left) and  \XeXe collisions at \snn = 5.44 TeV (right) with calculations using the HIJING and AMPT models. Lower panels: Ratios of the model calculations to measured $\sqrt{ \llangle  \Delta p_{\rm T1}\Delta p_{\rm T2} \rrangle }/{\llangle p_{\rm T} \rrangle }$.
    Solid symbols represent the measured data with statistical (vertical bars)  and systematic (boxes) uncertainties. Model calculations are shown with shaded bands denoting their statistical uncertainty. 
    }
    \label{fig:2}
\end{figure}

Figure~\ref{fig:2} shows the comparison of the evolution of the measured $\sqrt{ \llangle  \Delta p_{\rm T1}\Delta p_{\rm T2} \rrangle }/{\llangle  p_{\rm T} \rrangle }$ as a function of $\langle {\rm d}N_{\rm{ch}}/{\rm d}\eta\rangle$ in Pb--Pb (left) and Xe--Xe (right) collisions with the results obtained using HIJING and AMPT models. The presented AMPT calculations 
were obtained for two scenarios: AMPT default and AMPT string melting. \\

The bottom panels of Fig.~\ref{fig:2} show ratios of HIJING and AMPT calculations to the measured data. The observed magnitude of the normalized transverse momentum correlator, $\sqrt{\llangle  \Delta p_{{\rm T}1}\Delta p_{{\rm T}2} \rrangle }$$/\llangle  p_{\rm T} \rrangle$, obtained with the HIJING model exhibits a simple power-law dependence. This behavior is accurately depicted by a fit of $\sqrt{ \llangle  \Delta p_{{\rm T}1}\Delta p_{{\rm T}2} \rrangle }$$/\llangle  p_{\rm T} \rrangle $ \ $\propto \langle {\rm d}N_{\rm{ch}}/{\rm d}\eta\rangle^{\alpha}$ where the exponent $\alpha$ is determined to be   $\minus 0.504 \pm 0.007$ within the charged particle multiplicity density interval of $20 < \langle \rm{d}N_{\rm ch}/d\eta \rangle< 2500$.

\begin{figure}[ht!]
    \begin{center}
    \includegraphics[width = 0.49\textwidth]
       {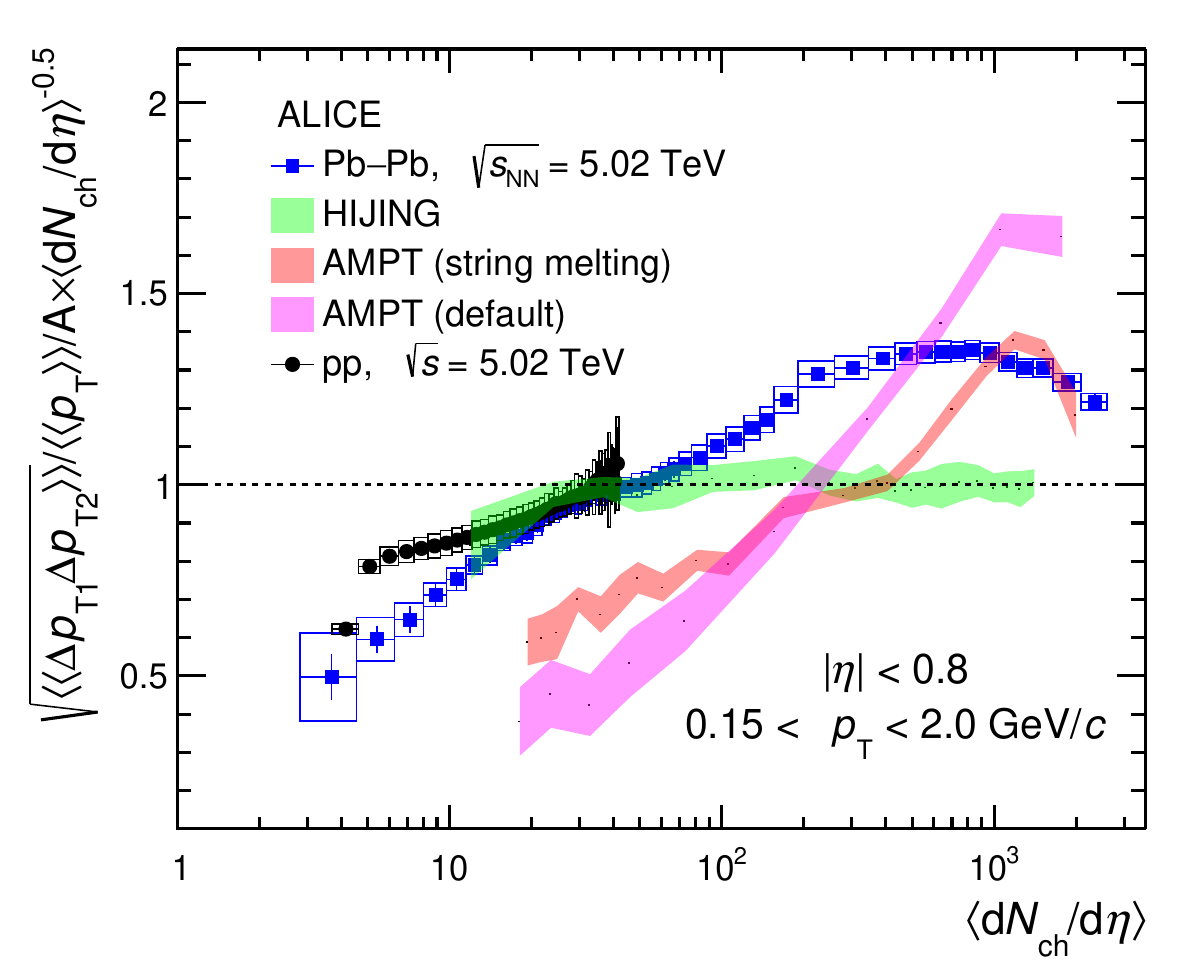}
      \includegraphics[width = 0.49\textwidth]
      {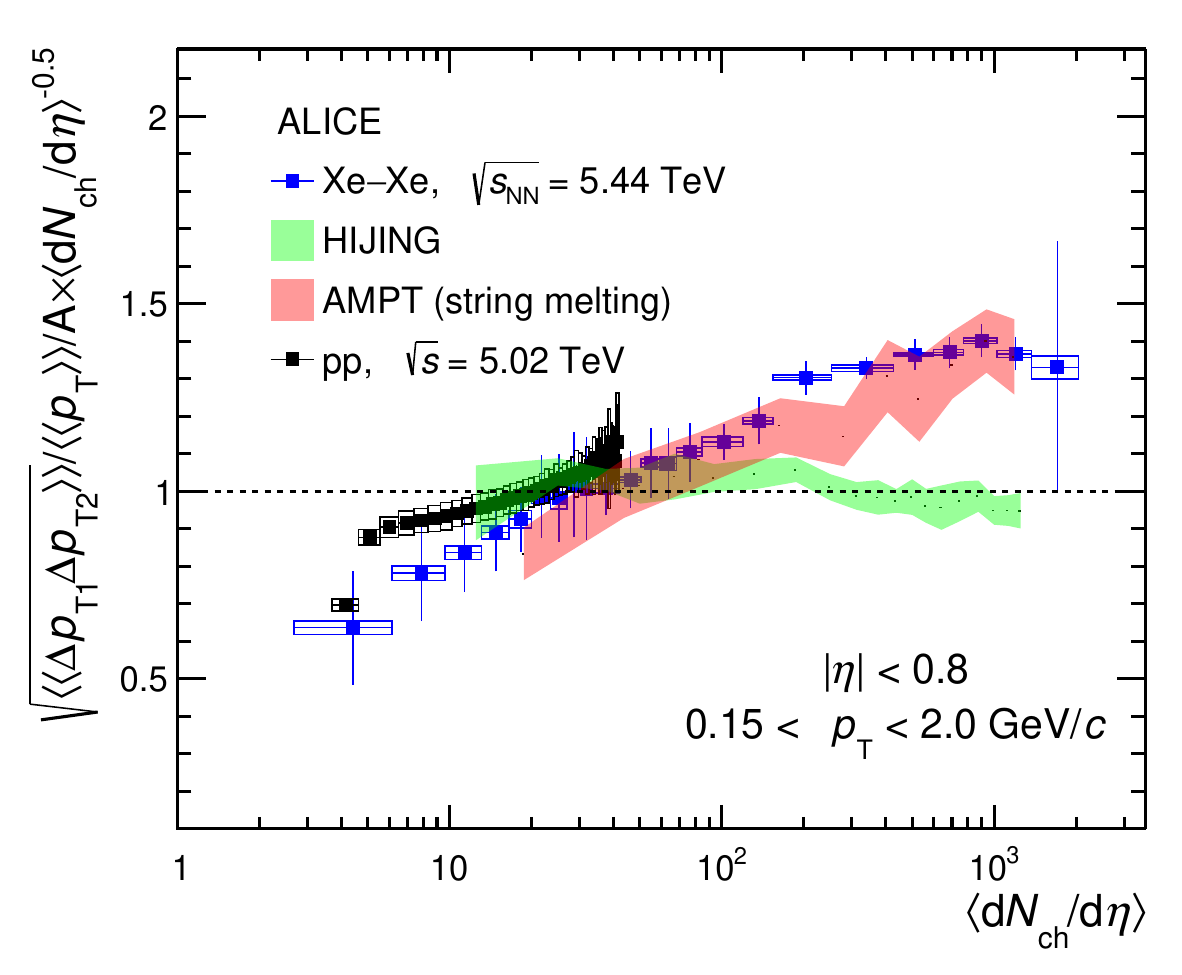}
    \end{center}
    \caption{Evolution of the ratio of $\sqrt{ \llangle  \Delta p_{\rm T1}\Delta p_{\rm T2} \rrangle }/{\llangle p_{\rm T} \rrangle }$ \ to results of a power-law fit to HIJING calculations: (left) \pp and \PbPb collisions at \snn = 5.02 TeV; (right) \pp collisions at \snn = 5.02 TeV and \XeXe collisions at \snn = 5.44 TeV. The power-law fit was performed on correlator values obtained with HIJING for \PbPb\ collisions at  \snn = 5.02 TeV and \XeXe collisions at \snn = 5.44 TeV for left and right panel, respectively, as described in the text. Solid symbols represent the measured data reported in this work with statistical (vertical bars)  and systematic (boxes) uncertainties. Calculations of the ratios obtained with HIJING and AMPT calculations are shown with shaded bands denoting   their statistical uncertainty.  }
    \label{fig:3}
\end{figure}

This power-law dependence and the exponent value are consistent with the behavior expected for a system consisting of a simple superposition of nucleon--nucleon collisions without rescattering of the secondaries as modeled by HIJING. One finds, however, that while the evolution of the correlator measured  in 
both \PbPb\ and \XeXe\ approximately follows the HIJING power-law fit in the low-multiplicity range $10< \langle {\rm d}N_{\rm{ch}}/{\rm d}\eta\rangle<50$, it clearly deviates from this simple trend at $\langle {\rm d}N_{\rm{ch}}/{\rm d}\eta\rangle>50$. This is also highlighted in Fig.~\ref{fig:3}~that shows the ratio of the magnitude of the $\sqrt{ \llangle  \Delta p_{{\rm T}1}\Delta p_{{\rm T}2} \rrangle }$$/\llangle  p_{\rm T} \rrangle $ correlator in \PbPb collisions at \snn = 5.02 TeV (left) and in \XeXe collisions at \snn = 5.44 TeV (right) to the power-law fit of \mpt fluctuations of HIJING model. This indicates that the final-state particle production in 
\PbPb collisions at \snn = 5.02 TeV and \XeXe at \snn = 5.44 TeV cannot be described by a mere superposition of independent particle-emitting sources. It also corroborates the earlier findings by the ALICE Collaboration  in \PbPb collisions at \snn = 2.76 TeV~\cite{ALICE:2014gvd} and those of the STAR Collaboration at RHIC energies~\cite{STAR:2003cbv,STAR:2005vxr,STAR:2019dow}. 

Deviations from a superposition model of independent particle-emitting sources, in A--A collisions,  are known to arise in measurements of nuclear modification factor~\cite{ALICE:2018vuu}, and anisotropic flow~\cite{ALICE:2011ab}, and other measurements of two-particle correlation functions~\cite{ALICE:2017xzf,ALICE:2019smr}. It is  thus reasonable to  seek theoretical guidance from a model  such as AMPT, which has had 
relative success in the description of data obtained at RHIC and LHC. One finds that the two versions of AMPT considered strongly under-predict the strength of the $\sqrt{ \llangle  \Delta p_{{\rm T}1}\Delta p_{{\rm T}2} \rrangle }$$/\llangle  p_{\rm T} \rrangle $\ correlator in the most peripheral Pb--Pb collisions. 
AMPT operated with string melting generates stronger collective flow because it effectively produces a dense partonic medium where interactions among quarks and gluons contribute significantly to the build-up of flow. Simulations with the AMPT model, with string melting, are in reasonable agreement with the data for \XeXe, and \PbPb collisions in the high-multiplicity regions. However, it is evident that both HIJING and AMPT lack some important features that determine the strength and evolution of $\sqrt{ \llangle  \Delta p_{{\rm T}1}\Delta p_{{\rm T}2} \rrangle }$$/\llangle  p_{\rm T} \rrangle $ \  with collision centrality for Pb--Pb collisions. For \XeXe collisions, AMPT with string melting on, shows a better agreement with data than for \PbPb collisions for low  multiplicity region. 
\begin{figure}[ht!]
    \begin{center}
    \includegraphics[width = 0.5\textwidth]
    {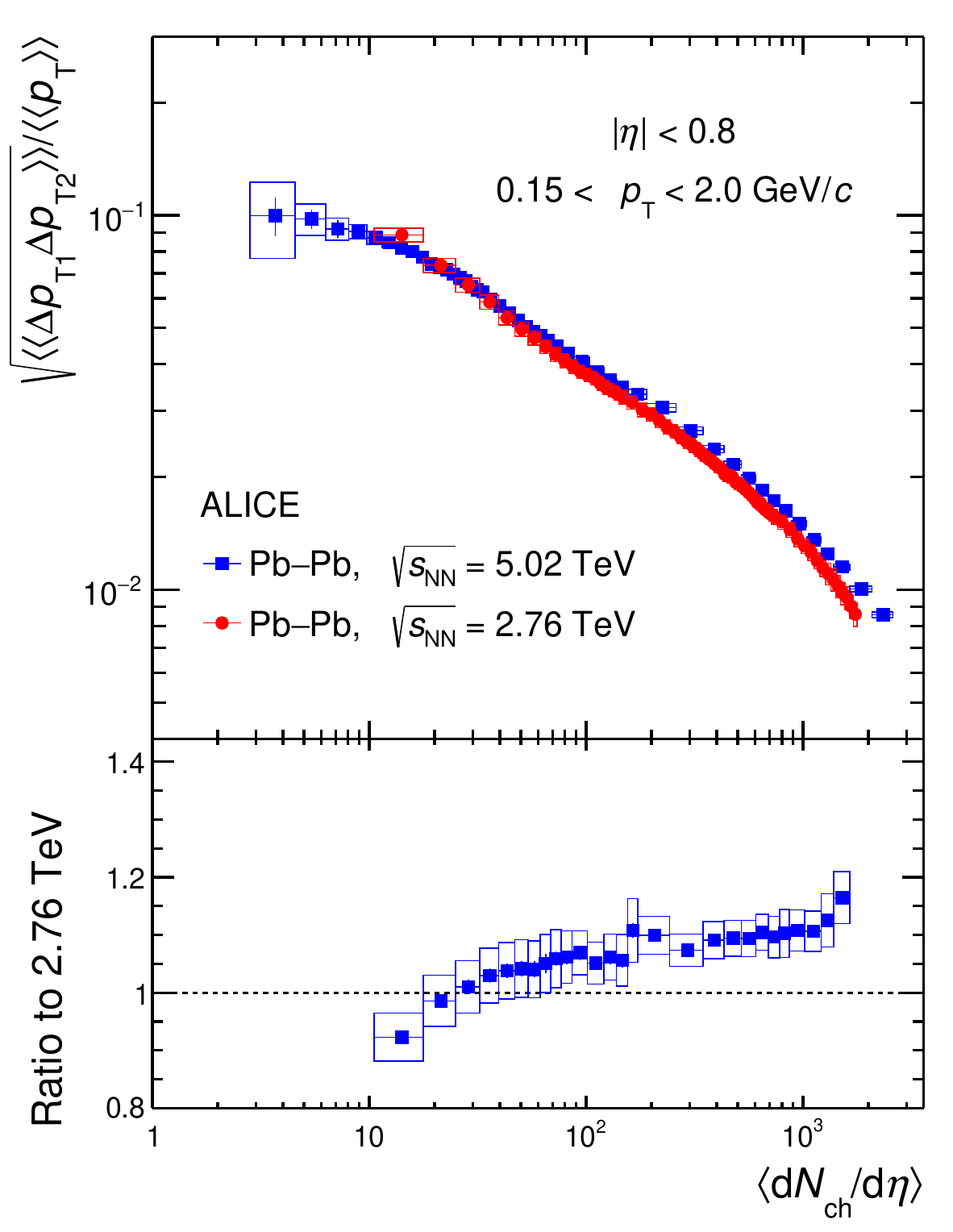}
    \end{center}
    \caption{(top) Normalized transverse momentum correlator, $\sqrt{ \llangle  \Delta p_{\rm T1}\Delta p_{\rm T2} \rrangle }/{\llangle p_{\rm T} \rrangle }$,  
    shown as a function of the charged-particle multiplicity density, $\langle{\rm d}N_{\rm ch}/{\rm d}\eta \rangle$, in \PbPb collisions at \snn = 2.76~\cite{ALICE:2014gvd} and 5.02 TeV; (bottom) Ratio of values of $\sqrt{ \llangle  \Delta p_{\rm T1}\Delta p_{\rm T2} \rrangle }/{\llangle p_{\rm T} \rrangle }$ measured at \snn = 5.02 TeV  to the corresponding results at \snn = 2.76 TeV. The statistical and systematic  uncertainties for both energies are represented by vertical bars and boxes, respectively.}
    \label{fig:4}
\end{figure}

In order to guide further theoretical inquiries, the strength and evolution of $\sqrt{ \llangle  \Delta p_{{\rm T}1}\Delta p_{{\rm T}2} \rrangle }$$/\llangle  p_{\rm T} \rrangle $ observed in pp, \PbPb\ and \XeXe\ collisions are compared in 
more detail. 
\begin{figure}[ht!]
    \begin{center}
    \includegraphics[width = 0.45\textwidth]
    {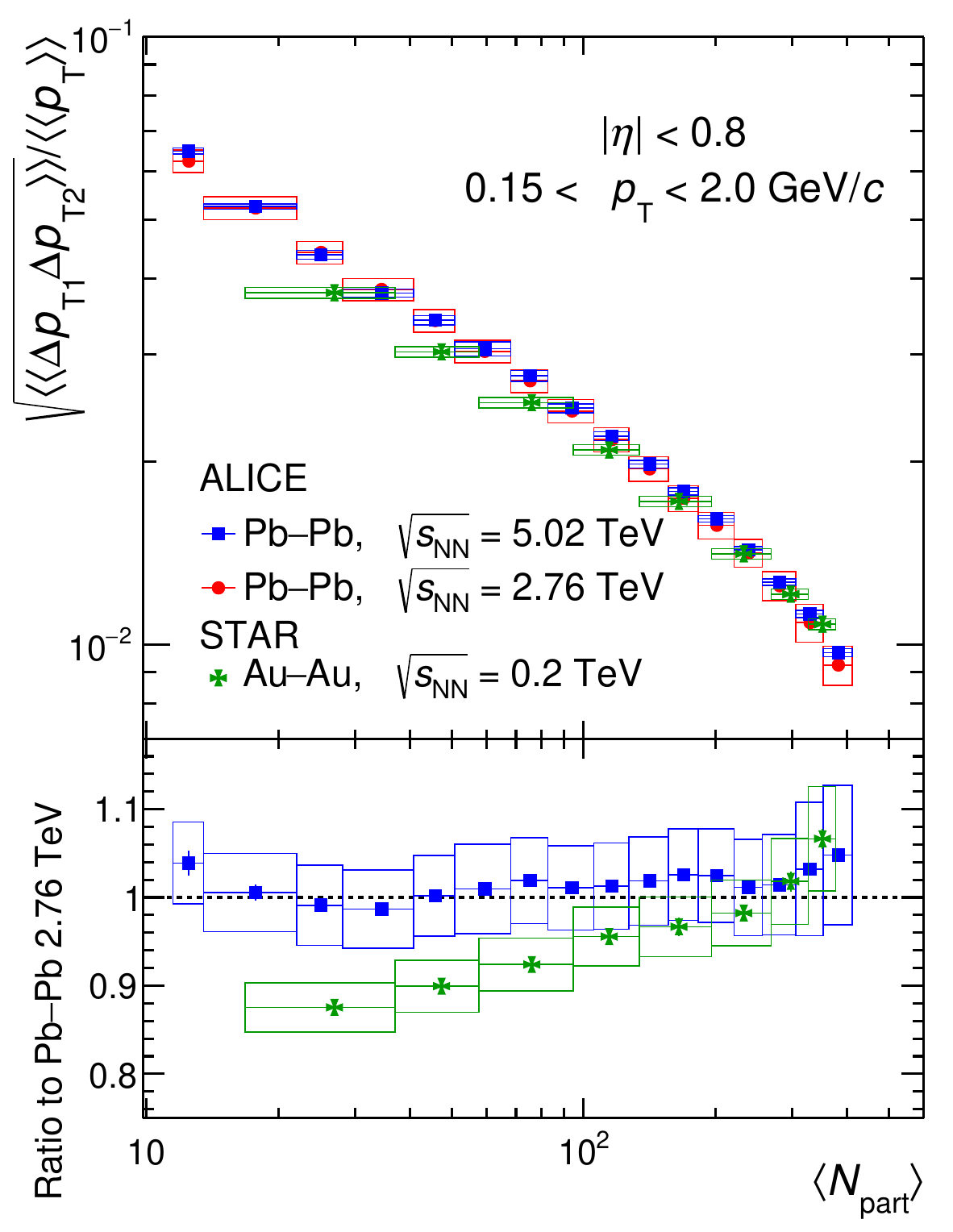}
    \includegraphics[width = 0.45\textwidth]
    {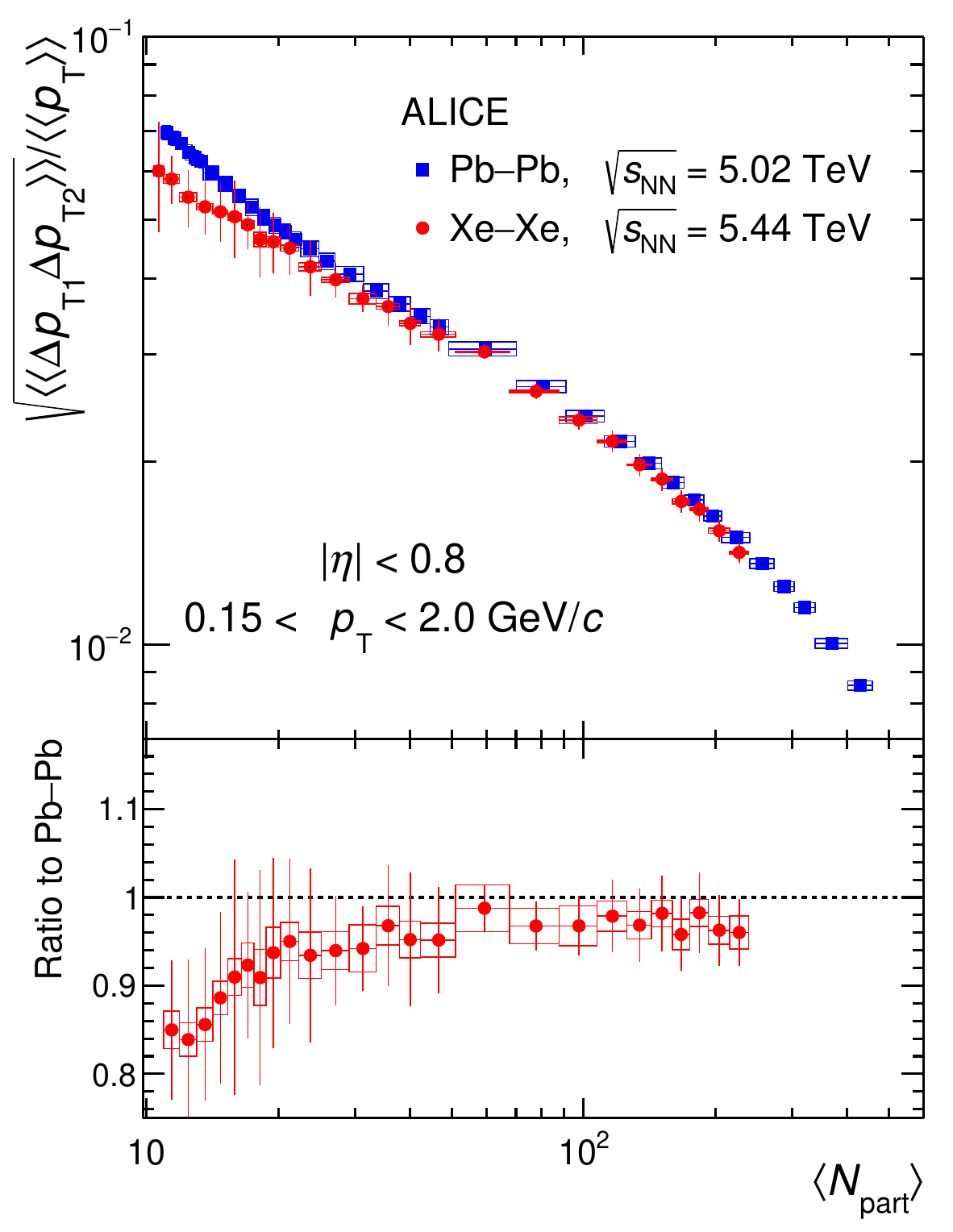}
    \end{center}
    \caption{ Left:  Normalized transverse momentum correlator, $\sqrt{ \llangle  \Delta p_{{\rm T}1}\Delta p_{{\rm T}2} \rrangle }/\llangle  p_{\rm T} \rrangle $,  
    shown as a function of the average number of participant nucleons, $\langle{ N_{\rm part}}\rangle$, in \PbPb collisions at \snn = 2.76 and 5.02 TeV and Au--Au collisions at 0.2 TeV. right: Normalized transverse momentum correlator, $\sqrt{ \llangle  \Delta p_{{\rm T}1}\Delta p_{{\rm T}2} \rrangle }/\llangle  p_{\rm T} \rrangle $,  
    plotted as a function $\langle{ N_{\rm part}} \rangle$ in \XeXe collisions at \snn = 5.44 TeV and \PbPb collisions at \snn = 5.02 TeV. The statistical and systematic  uncertainties are represented by vertical bars and boxes respectively.}
    \label{fig:5}
\end{figure}

Figure~\ref{fig:3} shows the ratio of measured values of $\sqrt{ \llangle  \Delta p_{{\rm T}1}\Delta p_{{\rm T}2} \rrangle }$$/\llangle  p_{\rm T} \rrangle $ \ to the HIJING results estimated by a power-law fit to the correlator values. Results are drawn as a function of the charged-particle multiplicity, $\langle {\rm d}N_{\rm{ch}}/{\rm d}\eta\rangle$.
The fitting procedure was carried over the range 25$<\langle {\rm d}N_{\rm{ch}}/{\rm d}\eta\rangle<$2500 with a fixed exponent value of $\alpha=-$0.5. The shaded green band is approximately centered at unity and shows that the power-law fit is a good description of the evolution of the strength of $\sqrt{ \llangle  \Delta p_{{\rm T}1}\Delta p_{{\rm T}2} \rrangle }$$/\llangle  p_{\rm T} \rrangle $ \ with $\langle {\rm d}N_{\rm{ch}}/{\rm d}\eta\rangle$ predicted by HIJING. Indeed, HIJING produces a progressive dilution of the correlator with rising values of $\langle {\rm d}N_{\rm{ch}}/{\rm d}\eta\rangle$ as expected.  By contrast, one finds that correlations observed in \PbPb\ and \XeXe\  increasingly undershoot the power-law fit at small densities while they significantly exceed the fit 
at $\langle {\rm d}N_{\rm{ch}}/{\rm d}\eta\rangle$ above 70--80, thereby signaling a considerable departure from
a system consisting of a superposition of independent nucleon--nucleon collisions. One additionally finds that both AMPT calculations considerably violate the density scaling. However, measurements in pp are well matched by HIJING simulations at multiplicities greater than 10.  

\subsection{Energy dependence of event-by-event 
\texorpdfstring{$\langle p_{\rm{T} }\rangle$}\xspace~ 
fluctuations}
\label{sec:energy}

Figure~\ref{fig:4} shows the dependence of the correlator strength $\sqrt{ \llangle  \Delta p_{{\rm T}1}\Delta p_{{\rm T}2} \rrangle }$$/\llangle  p_{\rm T} \rrangle $ as a function of charged-particle multiplicity ($\langle {\rm d}N_{\rm{ch}}/{\rm d}\eta\rangle$) for Pb--Pb collisions at two energies, $\sqrt{s_{\rm NN}}=$ 2.76 TeV and 5.02 TeV. The lower panel of the figure shows the ratio of 5.02 TeV correlator to that of 2.76 TeV. The ratio is close to unity at low values of $\langle {\rm d}N_{\rm{ch}}/{\rm d}\eta\rangle$ and increases with increasing multiplicity. For central collisions (large values of $\langle {\rm d}N_{\rm{ch}}/{\rm d}\eta\rangle$), the correlator at 5.02 TeV is up to 20\% larger compared to the one at 2.76 TeV.

Fig.~\ref{fig:5} (left panel) shows the dependence of the correlator as a function of the number of participant nucleons ($ N_{\rm part} $) for Pb--Pb collisions at two LHC energies and Au--Au collisions at $\sqrt{s_{\rm NN}}=$ 200 GeV at RHIC~\cite{STAR:2005vxr}. As shown in the ratio plot, the collision energy dependence in Pb--Pb collisions disappears when the correlator is shown as a function ($\langle N_{\rm part} \rangle$). This suggests that for a given initial-state overlap geometry, the correlator strength is independent of the collision energy. However, an energy dependence is observed when comparing RHIC and LHC energies.  The behavior in Fig.~\ref{fig:4} could be due to the dependence of the $\langle p_{\rm T} \rangle$ fluctuations on the collision energy or could be due to the larger number of particles produced at $\sqrt{s_{\rm NN}}=$ 5.02 TeV that shifts $\langle {\rm d}N_{\rm{ch}}/{\rm d}\eta\rangle$ to a higher value as compared to $\sqrt{s_{\rm NN}}=$ 2.76 TeV. The energy dependence as a function of $\langle N_{\rm part} \rangle$ in Fig.~\ref{fig:5} corroborates with the above statement. The right panel of Fig.~\ref{fig:5} shows that  similar values of $\sqrt{ \llangle  \Delta p_{{\rm T}1}\Delta p_{{\rm T}2} \rrangle }/\llangle  p_{\rm T} \rrangle $ are observed in  \PbPb and \XeXe collisions for $\langle N_{\rm part} \rangle > $ 25 but differ by as much as 15\% at smaller $\langle N_{\rm part} \rangle$ (i.e. very peripheral collisions). One can infer that the observed difference is  caused by the intrinsic deformation of the Xe nucleus. Indeed, evidence for the role of nuclear deformation on correlation observables has been reported in recent studies~\cite{ATLAS:2022dov}. However, reaching a definitive conclusion on the role of geometry fluctuations associated with the deformed Xe nucleus likely requires  detailed simulations of the strength of $\sqrt{ \llangle  \Delta p_{\rm T1}\Delta p_{\rm T2} \rrangle }/{\llangle p_{\rm T} \rrangle }$ with and without the inclusion of deformation. Such simulations were deemed beyond the scope of this work but will be considered in the future.

\subsection{Transverse spherocity dependence of \texorpdfstring{$\langle p_{\rm{T} }\rangle$}{average pt} fluctuations in pp collisions}
\label{sec:resultsVsSo}
\begin{figure}[ht!]
    \begin{center}
    \includegraphics[width = 0.5\textwidth]{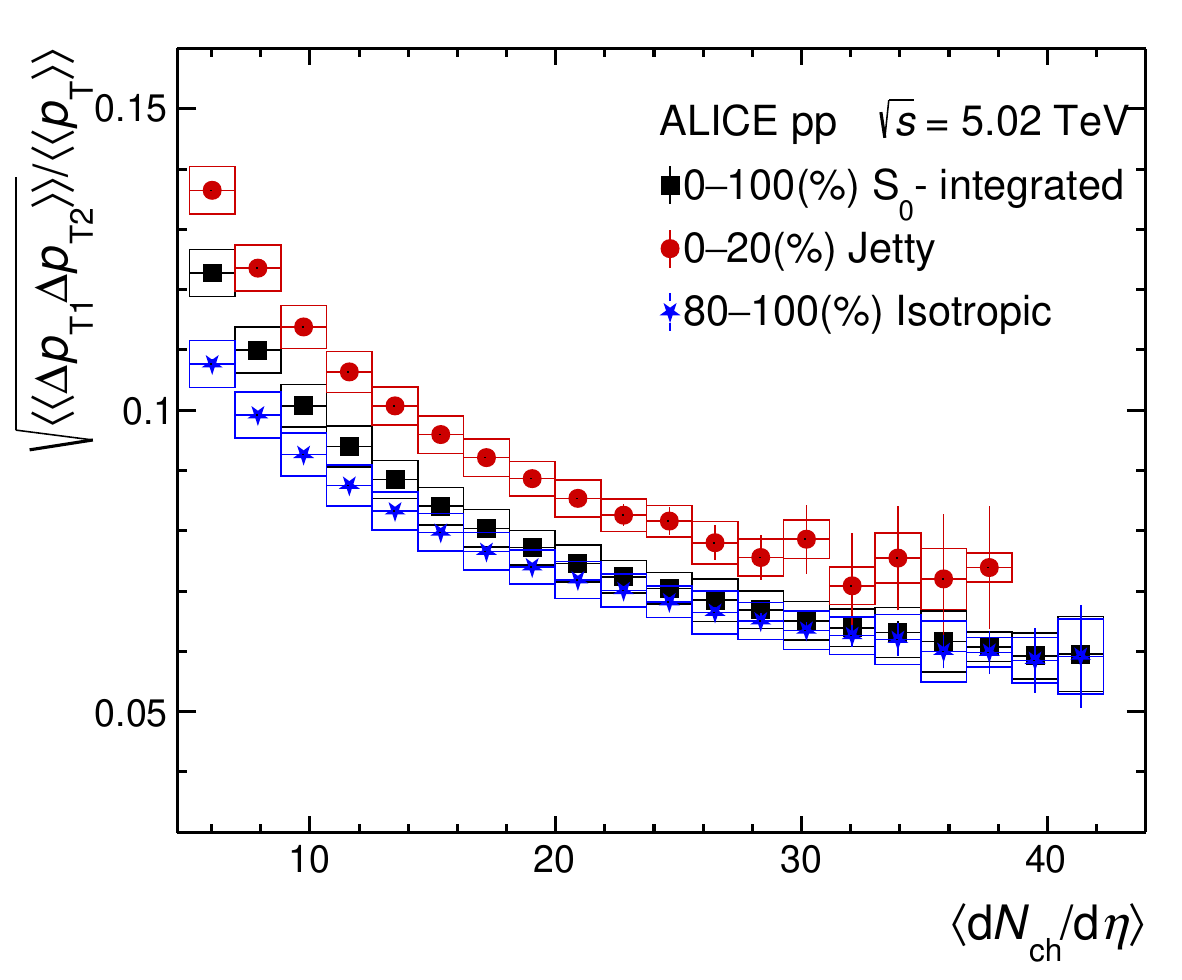}   
    \end{center}
    \caption{
    Comparison of the normalized transverse momentum correlator, $\sqrt{ \llangle  \Delta p_{{\rm T}1}\Delta p_{{\rm T}2} \rrangle }/\llangle  p_{\rm T} \rrangle $ as a function of the charged-particle multiplicity density for spherocity-integrated (black), jetty (red), and isotropic (blue) events in pp collisions at $\sqrt{s}=5.02$ TeV. The statistical and systematic uncertainties of the measured data for all spherocity classes are represented by vertical bars and boxes, respectively.}
    \label{fig:6}
\end{figure}

The results presented in Fig.~\ref{fig:3} indicate that the strength of the correlator in central Pb--Pb and Xe--Xe collisions deviates by as much as 30\% from the trivial scaling expected in the dilution scenario of independent particle-emitting sources. Although this deviation might stem largely from the kinematic focusing of correlated pairs associated with radial flow, it is of interest to examine whether the enhanced correlation values might arise from fluctuations associated with jet production and, more particularly, event-by-event variations in the number of jets and their composition. One might naively expect that fluctuations in the number or constituents of jets relative to ``baseline" collisions (with no jet) could increase
fluctuations and thus change the magnitude of the correlator. It is thus interesting to investigate, based on pp collisions alone, how the magnitude of the correlator changes from collisions where effects of jets are less pronounced to those featuring prominent jets. This investigation is conducted by studying the magnitude of the $\sqrt{ \llangle  \Delta p_{{\rm T}1}\Delta p_{{\rm T}2} \rrangle }/\llangle  p_{\rm T} \rrangle $ correlator relative to the shape of particle emission in the transverse plane. The spherocity observable, described in Sec.~\ref{sec:methodolgy}, is used to select collisions based on their transverse shape. Collisions with $S_{0}=1$ are expected to feature few or no jets, as by construction these are isotropic events without any preferred direction. This can be also possible for events with multijets. Whereas, events with $S_{0}\sim 0$ are expected to feature back-to-back jets. It is then nominally possible to get insight into the impact of jets on the magnitude of   $\sqrt{ \llangle  \Delta p_{{\rm T}1}\Delta p_{{\rm T}2} \rrangle }/\llangle  p_{\rm T} \rrangle $.
\begin{figure}[ht!]
    \begin{center}
    \includegraphics[width = 0.8\textwidth]{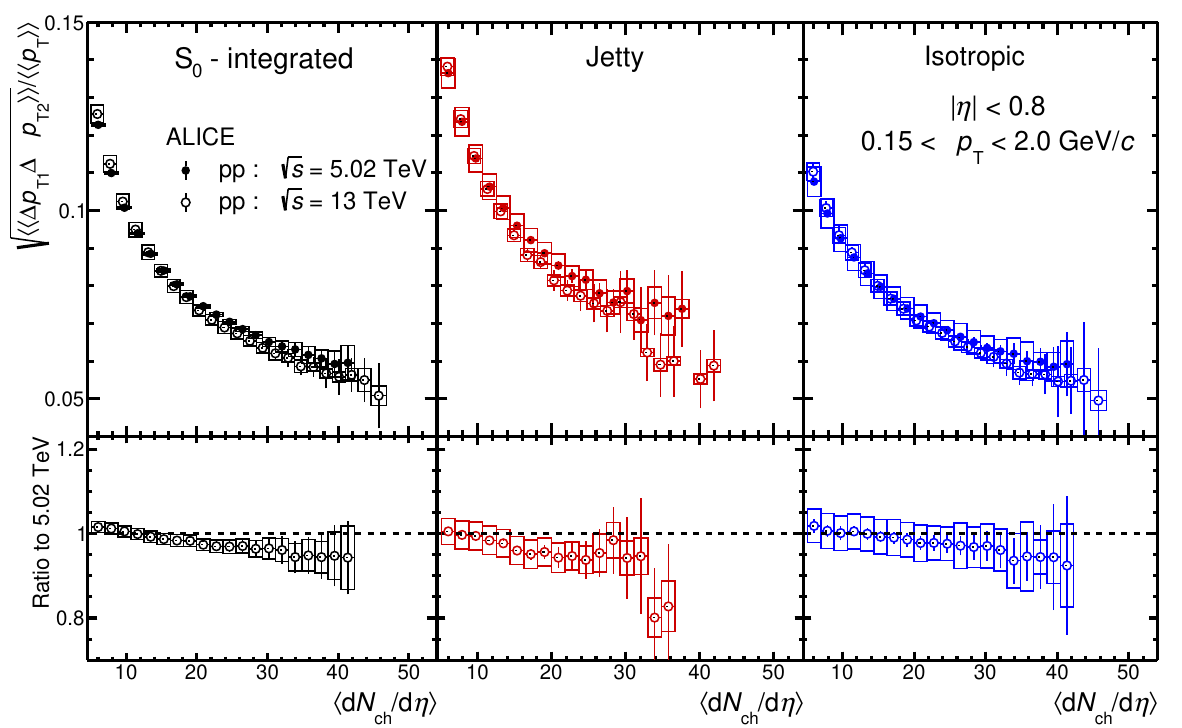}   
    \end{center}
    \caption{Upper panels: Comparison of the normalized transverse momentum correlator, $\sqrt{ \llangle  \Delta p_{{\rm T}1}\Delta p_{{\rm T}2} \rrangle }/\llangle  p_{\rm T} \rrangle $ as a function of charged-particle density in pp collisions at $\sqrt{s}=5.02$ and  \s = 13~TeV for spherocity-integrated (left), jetty (middle) and isotropic (right) events; Lower panels: Ratio of the $\sqrt{ \llangle  \Delta p_{{\rm T}1}\Delta p_{{\rm T}2} \rrangle }/\llangle  p_{\rm T} \rrangle $ in pp collisions at $\sqrt{s}=13$ TeV to $\sqrt{s}=5.02$ TeV. The statistical and systematic uncertainties of the measured data for all spherocity classes are represented by vertical bars and boxes, respectively.}
    \label{fig:7}
\end{figure}

Figure~\ref{fig:6} presents measurements of the evolution of the strength of the correlator $\sqrt{ \llangle  \Delta p_{{\rm T}1}\Delta p_{{\rm T}2} \rrangle }$$/\llangle  p_{\rm T} \rrangle $ \ as a function of $\langle {\rm d}N_{\rm{ch}}/{\rm d}\eta\rangle$ in pp collisions at \s = 5.02 TeV for selected spherocity classes. Black squares display inclusive events, i.e., events with no spherocity selection; red circles present events characterized by a back-to-back jet topology denoted as “jetty” in the following, with $S_{0} < 0.425$ corresponding to the 20\% of events with lower spherocity; and blue stars are for the 20\% most isotropic events, with $S_{0} > 0.745$. Low-spherocity events feature
$\sqrt{ \llangle  \Delta p_{{\rm T}1}\Delta p_{{\rm T}2} \rrangle }$$/\llangle  p_{\rm T} \rrangle $ values larger than those observed for isotropic events. 
The presence of a pronounced back-to-back jet topology is found to enhance the magnitude of the correlator by about 20\% in high-multiplicity collisions. This enhancement likely results from jet particles being emitted in a narrow cone and thus being more correlated on average than other particles. 
 \begin{figure}[ht!]
    \begin{center}
   \includegraphics[width = 0.9\textwidth]{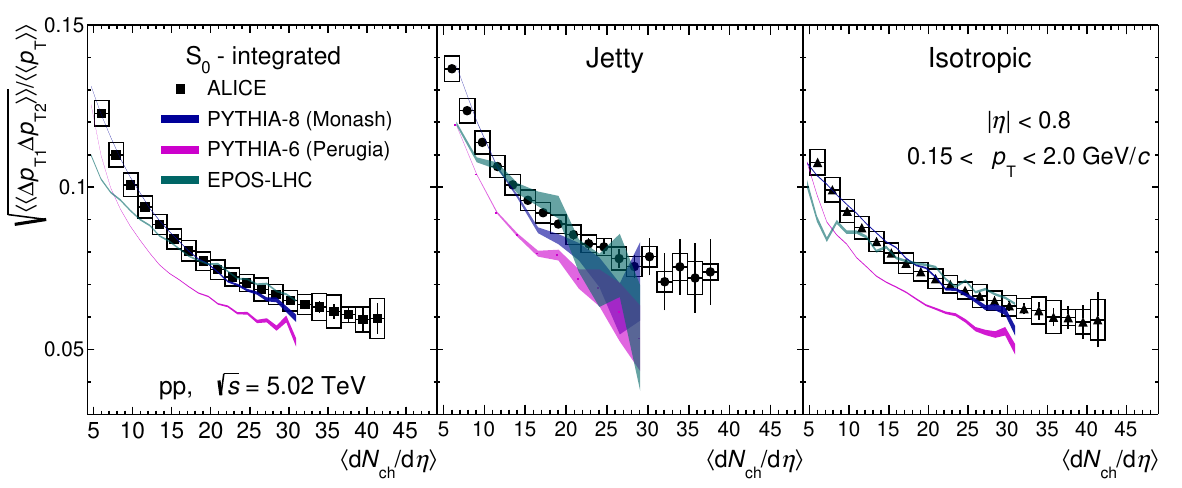} 
    \end{center}
    \caption{Comparison of normalized transverse momentum correlator,  $\sqrt{\llangle  \Delta p_{{\rm T}1}\Delta p_{{\rm T}2} \rrangle }/\llangle p_{\rm T} \rrangle $  as a function of the produced particle multiplicity for integrated (left), jetty (middle), and isotropic (right) pp collisions at \s = 5.02 TeV with calculations from the PYTHIA and EPOS models. See text for details. The statistical and systematic  uncertainties of the measured data for all spherocity classes are represented by vertical bars and boxes, respectively.} 
    \label{fig:8}
\end{figure}
It is interesting to consider whether the observed 20\% correlation strength difference between collisions dominated by a back-to-back jet topology and isotropic events may entail for correlation in larger systems. Since high-\pt particles and jets are suppressed in mid to central heavy-ion collisions, one expects that contributions to the $\sqrt{ \llangle  \Delta p_{\rm T1}\Delta p_{\rm T2} \rrangle }/{\llangle p_{\rm T} \rrangle }$ correlator, from jet particles, might also be suppressed. This should then reduce the correlation strength relative to the scaled dependence resulting from the 
dilution of the correlator in mid to central heavy-ion collisions. However, note that an excess is observed relative to the scaled dependence in both Xe--Xe and Pb--Pb collisions. One can then infer that  
the excess of correlation strength observed in these systems is not likely linked to jets but  rather arises from other causes. A possible cause of the increased strength may be the strong transverse radial flow arising from the rapid expansion of the matter formed in heavy-ion collisions~\cite{ALICE:2013mez,Melo:2015iva,Voloshin:2005qj}. Indeed, one expects that  the transverse radial expansion should accelerate correlated particles resulting from resonance (high mass hadrons) decays, string fragmentation, or QGP hadronization, giving thus rise to larger $\sqrt{ \llangle  \Delta p_{\rm T1}\Delta p_{\rm T2} \rrangle }/{\llangle p_{\rm T} \rrangle }$ correlator values. 

It is also interesting to consider how the magnitude of $\sqrt{ \llangle  \Delta p_{\rm T1}\Delta p_{\rm T2} \rrangle }/{\llangle p_{\rm T} \rrangle }$ evolves with $\sqrt{s}$ in pp collisions. 
Figure~\ref{fig:7} compares the dependence  of  $\sqrt{ \llangle  \Delta p_{\rm T1}\Delta p_{\rm T2} \rrangle }/{\llangle p_{\rm T} \rrangle }$ on $\langle {\rm d}N_{\rm{ch}}/{\rm d}\eta\rangle$ for inclusive, low- and high-spherocity events measured in \pp collisions at \s = 5.02 TeV and 13 TeV. One observes the correlation strength exhibits 
only a small dependence, if any, on the energy of the pp collisions
in both jetty and isotropic events. This indicates that changes in the strength of  $\sqrt{ \llangle  \Delta p_{\rm T1}\Delta p_{\rm T2} \rrangle }$ are essentially compensated by the rise in $\llangle p_{\rm T} \rrangle$ associated with the increase in the collision energy.

 The measured \mpt fluctuations for different spherocity classes are   compared in Fig.~\ref{fig:8} with calculations performed with the MC generators PYTHIA 6~\cite{Sjostrand:2006za}, PYTHIA 8~\cite{Skands:2014pea}, and EPOS LHC~\cite{Pierog:2013ria}. 
 PYTHIA 6 significantly underestimates the magnitude of $\sqrt{ \llangle  \Delta p_{\rm T1}\Delta p_{\rm T2} \rrangle }/{\llangle p_{\rm T} \rrangle }$, while PYTHIA 8 reproduces the data rather well. This indicates that the MPI mechanism is crucial for the description of particle production in PYTHIA. In addition, EPOS LHC, a model with core-corona approach, also reproduces the data well in both spherocity classes. 

\section{Summary and conclusions}
\label{sec:summary}
Event-by-event $\langle p_{\rm{T}}\rangle$ fluctuations of  charged particles produced in Pb--Pb and Xe--Xe collisions at $\sqrt{s_{\rm{NN}}}$ = 5.02 and  5.44 TeV, respectively,  and in pp  collisions at $\sqrt{s}$ = 5.02 TeV are studied  based on the normalized $\sqrt{\llangle  \Delta p_{{\rm T}1}\Delta p_{{\rm T}2} \rrangle }/\llangle  p_{\rm T} \rrangle $ integral correlator. The correlation strength is measured as a function of produced charged-particle multiplicity in all three collision systems and as a function of the
spherocity of produced particles at midrapidity in pp collisions.  The correlator strength is positive, thus indicating that significant dynamical fluctuations are observed in heavy-ion collisions and corroborating prior measurements reported by the STAR and ALICE Collaborations~\cite{Voloshin:2005qj,Huovinen:2001cy}. The strength of $\sqrt{\llangle  \Delta p_{{\rm T}1}\Delta p_{{\rm T}2} \rrangle }/\llangle  p_{\rm T} \rrangle $ is also observed to monotonically decrease with increasing multiplicity in all measured systems, likely resulting in part from a dilution of the correlation strength associated with an increase in particle production.  The observed decrease with multiplicity, however, significantly deviates from a power scaling of the form $\langle {\rm d}N_{\rm{ch}}/{\rm d}\eta\rangle^{-1/2}$ expected for a source consisting of a superposition of independent nucleon--nucleon (or parton--parton) collisions with no re-scattering of the secondaries. 
A comparison of the correlation strength measured in low- and high-spherocity pp collisions shows that the selection of events characterized by a back-to-back jet topology yields a 20\% increase in the correlator strength relative to isotropic events. Given that the production of high \pt particles and jets is known to be quenched in mid to central A--A collisions, one concludes that the observed deviation from $\langle {\rm d}N_{\rm{ch}}/{\rm d}\eta\rangle^{-1/2}$ is likely not associated with a change in jet production but originates from other sources. A prime candidate for such a source is the large transverse radial flow arising in mid to central A--A collisions~\cite{ALICE:2013mez,Melo:2015iva}.



\newenvironment{acknowledgement}{\relax}{\relax}

\begin{acknowledgement}

\section*{Acknowledgements}
\input{fa_2024-09-18_Opt_C.tex}
\end{acknowledgement}

\bibliographystyle{utphys}   
\bibliography{bibliography}

\newpage
\appendix

%
%

\section{The ALICE Collaboration}
\label{app:collab}
\input{Alice_Authorlist_2024-09-18_Opt_C.tex}
\end{document}

%% file: commands.tex
%

\newcommand{\pp}           {pp\xspace}
\newcommand{\ppbar}        {\mbox{$\mathrm {p\overline{p}}$}\xspace}
\newcommand{\XeXe}         {\mbox{Xe--Xe}\xspace}
\newcommand{\PbPb}         {\mbox{Pb--Pb}\xspace}
\newcommand{\pA}           {\mbox{pA}\xspace}
\newcommand{\pPb}          {\mbox{p--Pb}\xspace}
\newcommand{\AuAu}         {\mbox{Au--Au}\xspace}
\newcommand{\dAu}          {\mbox{d--Au}\xspace}

\newcommand{\s}            {\ensuremath{\sqrt{s}}\xspace}
\newcommand{\snn}          {\ensuremath{\sqrt{s_{\mathrm{NN}}}}\xspace}
\newcommand{\pt}           {\ensuremath{p_{\rm T}}\xspace}
\newcommand{\mpt}       {$\langle p_{\mathrm{T}}\rangle$\xspace}
\newcommand{\ycms}         {\ensuremath{y_{\rm CMS}}\xspace}
\newcommand{\ylab}         {\ensuremath{y_{\rm lab}}\xspace}
\newcommand{\etarange}[1]  {\mbox{$\left | \eta \right |~<~#1$}}
\newcommand{\yrange}[1]    {\mbox{$\left | y \right |~<~#1$}}
\newcommand{\dndy}         {\ensuremath{\mathrm{d}N_\mathrm{ch}/\mathrm{d}y}\xspace}
\newcommand{\dndeta}       {\ensuremath{\mathrm{d}N_\mathrm{ch}/\mathrm{d}\eta}\xspace}
\newcommand{\avdndeta}     {\ensuremath{\langle\dndeta\rangle}\xspace}
\newcommand{\dNdy}         {\ensuremath{\mathrm{d}N_\mathrm{ch}/\mathrm{d}y}\xspace}
\newcommand{\Npart}        {\ensuremath{N_\mathrm{part}}\xspace}
\newcommand{\Ncoll}        {\ensuremath{N_\mathrm{coll}}\xspace}
\newcommand{\dEdx}         {\ensuremath{\textrm{d}E/\textrm{d}x}\xspace}
\newcommand{\RpPb}         {\ensuremath{R_{\rm pPb}}\xspace}
\newcommand{\sDptDpt}    {$\sqrt{ \langle \Delta p_{\rm Ti}\Delta p_{\rm Tj} \rangle}/{\langle\langle p_{\rm T} \rangle\rangle}$}
\newcommand{\nineH}        {$\sqrt{s}~=~0.9$~Te\kern-.1emV\xspace}
\newcommand{\seven}        {$\sqrt{s}~=~7$~Te\kern-.1emV\xspace}
\newcommand{\twoH}         {$\sqrt{s}~=~0.2$~Te\kern-.1emV\xspace}
\newcommand{\twosevensix}  {$\sqrt{s}~=~2.76$~Te\kern-.1emV\xspace}
\newcommand{\five}         {$\sqrt{s}~=~5.02$~Te\kern-.1emV\xspace}
\newcommand{\twosevensixnn}{$\sqrt{s_{\mathrm{NN}}}~=~2.76$~Te\kern-.1emV\xspace}
\newcommand{\fivenn}       {$\sqrt{s_{\mathrm{NN}}}~=~5.02$~Te\kern-.1emV\xspace}
\newcommand{\LT}           {L{\'e}vy-Tsallis\xspace}
\newcommand{\GeVc}         {Ge\kern-.1emV/$c$\xspace}
\newcommand{\MeVc}         {Me\kern-.1emV/$c$\xspace}
\newcommand{\TeV}          {Te\kern-.1emV\xspace}
\newcommand{\GeV}          {Ge\kern-.1emV\xspace}
\newcommand{\MeV}          {Me\kern-.1emV\xspace}
\newcommand{\GeVmass}      {Ge\kern-.2emV/$c^2$\xspace}
\newcommand{\MeVmass}      {Me\kern-.2emV/$c^2$\xspace}
\newcommand{\lumi}         {\ensuremath{\mathcal{L}}\xspace}

\newcommand{\ITS}          {\rm{ITS}\xspace}
\newcommand{\TOF}          {\rm{TOF}\xspace}
\newcommand{\ZDC}          {\rm{ZDC}\xspace}
\newcommand{\ZDCs}         {\rm{ZDCs}\xspace}
\newcommand{\ZNA}          {\rm{ZNA}\xspace}
\newcommand{\ZNC}          {\rm{ZNC}\xspace}
\newcommand{\SPD}          {\rm{SPD}\xspace}
\newcommand{\SDD}          {\rm{SDD}\xspace}
\newcommand{\SSD}          {\rm{SSD}\xspace}
\newcommand{\TPC}          {\rm{TPC}\xspace}
\newcommand{\TRD}          {\rm{TRD}\xspace}
\newcommand{\VZERO}        {\rm{V0}\xspace}
\newcommand{\VZEROA}       {\rm{V0A}\xspace}
\newcommand{\VZEROC}       {\rm{V0C}\xspace}
\newcommand{\Vdecay} 	   {\ensuremath{V^{0}}\xspace}

\newcommand{\ee}           {\ensuremath{e^{+}e^{-}}} 
\newcommand{\pip}          {\ensuremath{\pi^{+}}\xspace}
\newcommand{\pim}          {\ensuremath{\pi^{-}}\xspace}
\newcommand{\kap}          {\ensuremath{\rm{K}^{+}}\xspace}
\newcommand{\kam}          {\ensuremath{\rm{K}^{-}}\xspace}
\newcommand{\pbar}         {\ensuremath{\rm\overline{p}}\xspace}
\newcommand{\kzero}        {\ensuremath{{\rm K}^{0}_{\rm{S}}}\xspace}
\newcommand{\lmb}          {\ensuremath{\Lambda}\xspace}
\newcommand{\almb}         {\ensuremath{\overline{\Lambda}}\xspace}
\newcommand{\Om}           {\ensuremath{\Omega^-}\xspace}
\newcommand{\Mo}           {\ensuremath{\overline{\Omega}^+}\xspace}
\newcommand{\X}            {\ensuremath{\Xi^-}\xspace}
\newcommand{\Ix}           {\ensuremath{\overline{\Xi}^+}\xspace}
\newcommand{\Xis}          {\ensuremath{\Xi^{\pm}}\xspace}
\newcommand{\Oms}          {\ensuremath{\Omega^{\pm}}\xspace}
\newcommand{\degree}       {\ensuremath{^{\rm o}}\xspace}

%% file: fa_2024-09-18_Opt_C.tex

The ALICE Collaboration would like to thank all its engineers and technicians for their invaluable contributions to the construction of the experiment and the CERN accelerator teams for the outstanding performance of the LHC complex.
The ALICE Collaboration gratefully acknowledges the resources and support provided by all Grid centres and the Worldwide LHC Computing Grid (WLCG) collaboration.
The ALICE Collaboration acknowledges the following funding agencies for their support in building and running the ALICE detector:
A. I. Alikhanyan National Science Laboratory (Yerevan Physics Institute) Foundation (ANSL), State Committee of Science and World Federation of Scientists (WFS), Armenia;
Austrian Academy of Sciences, Austrian Science Fund (FWF): [M 2467-N36] and Nationalstiftung f\"{u}r Forschung, Technologie und Entwicklung, Austria;
Ministry of Communications and High Technologies, National Nuclear Research Center, Azerbaijan;
Conselho Nacional de Desenvolvimento Cient\'{\i}fico e Tecnol\'{o}gico (CNPq), Financiadora de Estudos e Projetos (Finep), Funda\c{c}\~{a}o de Amparo \`{a} Pesquisa do Estado de S\~{a}o Paulo (FAPESP) and Universidade Federal do Rio Grande do Sul (UFRGS), Brazil;
Bulgarian Ministry of Education and Science, within the National Roadmap for Research Infrastructures 2020-2027 (object CERN), Bulgaria;
Ministry of Education of China (MOEC) , Ministry of Science \& Technology of China (MSTC) and National Natural Science Foundation of China (NSFC), China;
Ministry of Science and Education and Croatian Science Foundation, Croatia;
Centro de Aplicaciones Tecnol\'{o}gicas y Desarrollo Nuclear (CEADEN), Cubaenerg\'{\i}a, Cuba;
Ministry of Education, Youth and Sports of the Czech Republic, Czech Republic;
The Danish Council for Independent Research | Natural Sciences, the VILLUM FONDEN and Danish National Research Foundation (DNRF), Denmark;
Helsinki Institute of Physics (HIP), Finland;
Commissariat \`{a} l'Energie Atomique (CEA) and Institut National de Physique Nucl\'{e}aire et de Physique des Particules (IN2P3) and Centre National de la Recherche Scientifique (CNRS), France;
Bundesministerium f\"{u}r Bildung und Forschung (BMBF) and GSI Helmholtzzentrum f\"{u}r Schwerionenforschung GmbH, Germany;
General Secretariat for Research and Technology, Ministry of Education, Research and Religions, Greece;
National Research, Development and Innovation Office, Hungary;
Department of Atomic Energy Government of India (DAE), Department of Science and Technology, Government of India (DST), University Grants Commission, Government of India (UGC) and Council of Scientific and Industrial Research (CSIR), India;
National Research and Innovation Agency - BRIN, Indonesia;
Istituto Nazionale di Fisica Nucleare (INFN), Italy;
Japanese Ministry of Education, Culture, Sports, Science and Technology (MEXT) and Japan Society for the Promotion of Science (JSPS) KAKENHI, Japan;
Consejo Nacional de Ciencia (CONACYT) y Tecnolog\'{i}a, through Fondo de Cooperaci\'{o}n Internacional en Ciencia y Tecnolog\'{i}a (FONCICYT) and Direcci\'{o}n General de Asuntos del Personal Academico (DGAPA), Mexico;
Nederlandse Organisatie voor Wetenschappelijk Onderzoek (NWO), Netherlands;
The Research Council of Norway, Norway;
Pontificia Universidad Cat\'{o}lica del Per\'{u}, Peru;
Ministry of Science and Higher Education, National Science Centre and WUT ID-UB, Poland;
Korea Institute of Science and Technology Information and National Research Foundation of Korea (NRF), Republic of Korea;
Ministry of Education and Scientific Research, Institute of Atomic Physics, Ministry of Research and Innovation and Institute of Atomic Physics and Universitatea Nationala de Stiinta si Tehnologie Politehnica Bucuresti, Romania;
Ministry of Education, Science, Research and Sport of the Slovak Republic, Slovakia;
National Research Foundation of South Africa, South Africa;
Swedish Research Council (VR) and Knut \& Alice Wallenberg Foundation (KAW), Sweden;
European Organization for Nuclear Research, Switzerland;
Suranaree University of Technology (SUT), National Science and Technology Development Agency (NSTDA) and National Science, Research and Innovation Fund (NSRF via PMU-B B05F650021), Thailand;
Turkish Energy, Nuclear and Mineral Research Agency (TENMAK), Turkey;
National Academy of  Sciences of Ukraine, Ukraine;
Science and Technology Facilities Council (STFC), United Kingdom;
National Science Foundation of the United States of America (NSF) and United States Department of Energy, Office of Nuclear Physics (DOE NP), United States of America.
In addition, individual groups or members have received support from:
Czech Science Foundation (grant no. 23-07499S), Czech Republic;
FORTE project, reg.\ no.\ CZ.02.01.01/00/22\_008/0004632, Czech Republic, co-funded by the European Union, Czech Republic;
European Research Council (grant no. 950692), European Union;
ICSC - Centro Nazionale di Ricerca in High Performance Computing, Big Data and Quantum Computing, European Union - NextGenerationEU;
Academy of Finland (Center of Excellence in Quark Matter) (grant nos. 346327, 346328), Finland;
Deutsche Forschungs Gemeinschaft (DFG, German Research Foundation) ``Neutrinos and Dark Matter in Astro- and Particle Physics'' (grant no. SFB 1258), Germany.

%% file: Alice_Authorlist_2024-09-18_Opt_C.tex
\begin{flushleft} 
\small

S.~Acharya\,\orcidlink{0000-0002-9213-5329}\,$^{\rm 126}$, 
A.~Agarwal$^{\rm 134}$, 
G.~Aglieri Rinella\,\orcidlink{0000-0002-9611-3696}\,$^{\rm 32}$, 
L.~Aglietta\,\orcidlink{0009-0003-0763-6802}\,$^{\rm 24}$, 
M.~Agnello\,\orcidlink{0000-0002-0760-5075}\,$^{\rm 29}$, 
N.~Agrawal\,\orcidlink{0000-0003-0348-9836}\,$^{\rm 25}$, 
Z.~Ahammed\,\orcidlink{0000-0001-5241-7412}\,$^{\rm 134}$, 
S.~Ahmad\,\orcidlink{0000-0003-0497-5705}\,$^{\rm 15}$, 
S.U.~Ahn\,\orcidlink{0000-0001-8847-489X}\,$^{\rm 71}$, 
I.~Ahuja\,\orcidlink{0000-0002-4417-1392}\,$^{\rm 36}$
% ...

\section*{Affiliation Notes}

$^{\rm I}$ Deceased\\
$^{\rm II}$ Also at: Max-Planck-Institut fur Physik, Munich, Germany\\
$^{\rm III}$ Also at: Italian National Agency for New Technologies, Energy and Sustainable Economic Development (ENEA), Bologna, Italy\\
$^{\rm IV}$ Also at: Dipartimento DET del Politecnico di Torino, Turin, Italy\\
$^{\rm V}$ Also at: Department of Applied Physics, Aligarh Muslim University, Aligarh, India\\
$^{\rm VI}$ Also at: Institute of Theoretical Physics, University of Wroclaw, Poland\\
$^{\rm VII}$ Also at: Facultad de Ciencias, Universidad Nacional Autónoma de México, Mexico City, Mexico\\

\section*{Collaboration Institutes}

$^{1}$ A.I. Alikhanyan National Science Laboratory (Yerevan Physics Institute) Foundation, Yerevan, Armenia\\
$^{2}$ AGH University of Krakow, Cracow, Poland\\
$^{3}$ Bogolyubov Institute for Theoretical Physics, National Academy of Sciences of Ukraine, Kiev, Ukraine\\
$^{4}$ Bose Institute, Department of Physics  and Centre for Astroparticle Physics and Space Science (CAPSS), Kolkata, India\\
$^{5}$ California Polytechnic State University, San Luis Obispo, California, United States\\
$^{6}$ Central China Normal University, Wuhan, China\\
$^{7}$ Centro de Aplicaciones Tecnol\'{o}gicas y Desarrollo Nuclear (CEADEN), Havana, Cuba\\
$^{8}$ Centro de Investigaci\'{o}n y de Estudios Avanzados (CINVESTAV), Mexico City and M\'{e}rida, Mexico\\
$^{9}$ Chicago State University, Chicago, Illinois, United States\\
$^{10}$ China Institute of Atomic Energy, Beijing, China\\
$^{11}$ China University of Geosciences, Wuhan, China\\
$^{12}$ Chungbuk National University, Cheongju, Republic of Korea\\
$^{13}$ Comenius University Bratislava, Faculty of Mathematics, Physics and Informatics, Bratislava, Slovak Republic\\
$^{14}$ Creighton University, Omaha, Nebraska, United States\\
$^{15}$ Department of Physics, Aligarh Muslim University, Aligarh, India\\
$^{16}$ Department of Physics, Pusan National University, Pusan, Republic of Korea\\
$^{17}$ Department of Physics, Sejong University, Seoul, Republic of Korea\\
$^{18}$ Department of Physics, University of California, Berkeley, California, United States\\
$^{19}$ Department of Physics, University of Oslo, Oslo, Norway\\
$^{20}$ Department of Physics and Technology, University of Bergen, Bergen, Norway\\
$^{21}$ Dipartimento di Fisica, Universit\`{a} di Pavia, Pavia, Italy\\
$^{22}$ Dipartimento di Fisica dell'Universit\`{a} and Sezione INFN, Cagliari, Italy\\
$^{23}$ Dipartimento di Fisica dell'Universit\`{a} and Sezione INFN, Trieste, Italy\\
$^{24}$ Dipartimento di Fisica dell'Universit\`{a} and Sezione INFN, Turin, Italy\\
$^{25}$ Dipartimento di Fisica e Astronomia dell'Universit\`{a} and Sezione INFN, Bologna, Italy\\
$^{26}$ Dipartimento di Fisica e Astronomia dell'Universit\`{a} and Sezione INFN, Catania, Italy\\
$^{27}$ Dipartimento di Fisica e Astronomia dell'Universit\`{a} and Sezione INFN, Padova, Italy\\
$^{28}$ Dipartimento di Fisica `E.R.~Caianiello' dell'Universit\`{a} and Gruppo Collegato INFN, Salerno, Italy\\
$^{29}$ Dipartimento DISAT del Politecnico and Sezione INFN, Turin, Italy\\
$^{30}$ Dipartimento di Scienze MIFT, Universit\`{a} di Messina, Messina, Italy\\
$^{31}$ Dipartimento Interateneo di Fisica `M.~Merlin' and Sezione INFN, Bari, Italy\\
$^{32}$ European Organization for Nuclear Research (CERN), Geneva, Switzerland\\
$^{33}$ Faculty of Electrical Engineering, Mechanical Engineering and Naval Architecture, University of Split, Split, Croatia\\
$^{34}$ Faculty of Nuclear Sciences and Physical Engineering, Czech Technical University in Prague, Prague, Czech Republic\\
$^{35}$ Faculty of Physics, Sofia University, Sofia, Bulgaria\\
$^{36}$ Faculty of Science, P.J.~\v{S}af\'{a}rik University, Ko\v{s}ice, Slovak Republic\\
$^{37}$ Faculty of Technology, Environmental and Social Sciences, Bergen, Norway\\
$^{38}$ Frankfurt Institute for Advanced Studies, Johann Wolfgang Goethe-Universit\"{a}t Frankfurt, Frankfurt, Germany\\
$^{39}$ Fudan University, Shanghai, China\\
$^{40}$ Gangneung-Wonju National University, Gangneung, Republic of Korea\\
$^{41}$ Gauhati University, Department of Physics, Guwahati, India\\
$^{42}$ Helmholtz-Institut f\"{u}r Strahlen- und Kernphysik, Rheinische Friedrich-Wilhelms-Universit\"{a}t Bonn, Bonn, Germany\\
$^{43}$ Helsinki Institute of Physics (HIP), Helsinki, Finland\\
$^{44}$ High Energy Physics Group,  Universidad Aut\'{o}noma de Puebla, Puebla, Mexico\\
$^{45}$ Horia Hulubei National Institute of Physics and Nuclear Engineering, Bucharest, Romania\\
$^{46}$ HUN-REN Wigner Research Centre for Physics, Budapest, Hungary\\
$^{47}$ Indian Institute of Technology Bombay (IIT), Mumbai, India\\
$^{48}$ Indian Institute of Technology Indore, Indore, India\\
$^{49}$ INFN, Laboratori Nazionali di Frascati, Frascati, Italy\\
$^{50}$ INFN, Sezione di Bari, Bari, Italy\\
$^{51}$ INFN, Sezione di Bologna, Bologna, Italy\\
$^{52}$ INFN, Sezione di Cagliari, Cagliari, Italy\\
$^{53}$ INFN, Sezione di Catania, Catania, Italy\\
$^{54}$ INFN, Sezione di Padova, Padova, Italy\\
$^{55}$ INFN, Sezione di Pavia, Pavia, Italy\\
$^{56}$ INFN, Sezione di Torino, Turin, Italy\\
$^{57}$ INFN, Sezione di Trieste, Trieste, Italy\\
$^{58}$ Inha University, Incheon, Republic of Korea\\
$^{59}$ Institute for Gravitational and Subatomic Physics (GRASP), Utrecht University/Nikhef, Utrecht, Netherlands\\
$^{60}$ Institute of Experimental Physics, Slovak Academy of Sciences, Ko\v{s}ice, Slovak Republic\\
$^{61}$ Institute of Physics, Homi Bhabha National Institute, Bhubaneswar, India\\
$^{62}$ Institute of Physics of the Czech Academy of Sciences, Prague, Czech Republic\\
$^{63}$ Institute of Space Science (ISS), Bucharest, Romania\\
$^{64}$ Institut f\"{u}r Kernphysik, Johann Wolfgang Goethe-Universit\"{a}t Frankfurt, Frankfurt, Germany\\
$^{65}$ Instituto de Ciencias Nucleares, Universidad Nacional Aut\'{o}noma de M\'{e}xico, Mexico City, Mexico\\
$^{66}$ Instituto de F\'{i}sica, Universidade Federal do Rio Grande do Sul (UFRGS), Porto Alegre, Brazil\\
$^{67}$ Instituto de F\'{\i}sica, Universidad Nacional Aut\'{o}noma de M\'{e}xico, Mexico City, Mexico\\
$^{68}$ iThemba LABS, National Research Foundation, Somerset West, South Africa\\
$^{69}$ Jeonbuk National University, Jeonju, Republic of Korea\\
$^{70}$ Johann-Wolfgang-Goethe Universit\"{a}t Frankfurt Institut f\"{u}r Informatik, Fachbereich Informatik und Mathematik, Frankfurt, Germany\\
$^{71}$ Korea Institute of Science and Technology Information, Daejeon, Republic of Korea\\
$^{72}$ Laboratoire de Physique Subatomique et de Cosmologie, Universit\'{e} Grenoble-Alpes, CNRS-IN2P3, Grenoble, France\\
$^{73}$ Lawrence Berkeley National Laboratory, Berkeley, California, United States\\
$^{74}$ Lund University Department of Physics, Division of Particle Physics, Lund, Sweden\\
$^{75}$ Nagasaki Institute of Applied Science, Nagasaki, Japan\\
$^{76}$ Nara Women{'}s University (NWU), Nara, Japan\\
$^{77}$ National and Kapodistrian University of Athens, School of Science, Department of Physics , Athens, Greece\\
$^{78}$ National Centre for Nuclear Research, Warsaw, Poland\\
$^{79}$ National Institute of Science Education and Research, Homi Bhabha National Institute, Jatni, India\\
$^{80}$ National Nuclear Research Center, Baku, Azerbaijan\\
$^{81}$ National Research and Innovation Agency - BRIN, Jakarta, Indonesia\\
$^{82}$ Niels Bohr Institute, University of Copenhagen, Copenhagen, Denmark\\
$^{83}$ Nikhef, National institute for subatomic physics, Amsterdam, Netherlands\\
$^{84}$ Nuclear Physics Group, STFC Daresbury Laboratory, Daresbury, United Kingdom\\
$^{85}$ Nuclear Physics Institute of the Czech Academy of Sciences, Husinec-\v{R}e\v{z}, Czech Republic\\
$^{86}$ Oak Ridge National Laboratory, Oak Ridge, Tennessee, United States\\
$^{87}$ Ohio State University, Columbus, Ohio, United States\\
$^{88}$ Physics department, Faculty of science, University of Zagreb, Zagreb, Croatia\\
$^{89}$ Physics Department, Panjab University, Chandigarh, India\\
$^{90}$ Physics Department, University of Jammu, Jammu, India\\
$^{91}$ Physics Program and International Institute for Sustainability with Knotted Chiral Meta Matter (WPI-SKCM$^{2}$), Hiroshima University, Hiroshima, Japan\\
$^{92}$ Physikalisches Institut, Eberhard-Karls-Universit\"{a}t T\"{u}bingen, T\"{u}bingen, Germany\\
$^{93}$ Physikalisches Institut, Ruprecht-Karls-Universit\"{a}t Heidelberg, Heidelberg, Germany\\
$^{94}$ Physik Department, Technische Universit\"{a}t M\"{u}nchen, Munich, Germany\\
$^{95}$ Politecnico di Bari and Sezione INFN, Bari, Italy\\
$^{96}$ Research Division and ExtreMe Matter Institute EMMI, GSI Helmholtzzentrum f\"ur Schwerionenforschung GmbH, Darmstadt, Germany\\
$^{97}$ Saga University, Saga, Japan\\
$^{98}$ Saha Institute of Nuclear Physics, Homi Bhabha National Institute, Kolkata, India\\
$^{99}$ School of Physics and Astronomy, University of Birmingham, Birmingham, United Kingdom\\
$^{100}$ Secci\'{o}n F\'{\i}sica, Departamento de Ciencias, Pontificia Universidad Cat\'{o}lica del Per\'{u}, Lima, Peru\\
$^{101}$ Stefan Meyer Institut f\"{u}r Subatomare Physik (SMI), Vienna, Austria\\
$^{102}$ SUBATECH, IMT Atlantique, Nantes Universit\'{e}, CNRS-IN2P3, Nantes, France\\
$^{103}$ Sungkyunkwan University, Suwon City, Republic of Korea\\
$^{104}$ Suranaree University of Technology, Nakhon Ratchasima, Thailand\\
$^{105}$ Technical University of Ko\v{s}ice, Ko\v{s}ice, Slovak Republic\\
$^{106}$ The Henryk Niewodniczanski Institute of Nuclear Physics, Polish Academy of Sciences, Cracow, Poland\\
$^{107}$ The University of Texas at Austin, Austin, Texas, United States\\
$^{108}$ Universidad Aut\'{o}noma de Sinaloa, Culiac\'{a}n, Mexico\\
$^{109}$ Universidade de S\~{a}o Paulo (USP), S\~{a}o Paulo, Brazil\\
$^{110}$ Universidade Estadual de Campinas (UNICAMP), Campinas, Brazil\\
$^{111}$ Universidade Federal do ABC, Santo Andre, Brazil\\
$^{112}$ Universitatea Nationala de Stiinta si Tehnologie Politehnica Bucuresti, Bucharest, Romania\\
$^{113}$ University of Cape Town, Cape Town, South Africa\\
$^{114}$ University of Derby, Derby, United Kingdom\\
$^{115}$ University of Houston, Houston, Texas, United States\\
$^{116}$ University of Jyv\"{a}skyl\"{a}, Jyv\"{a}skyl\"{a}, Finland\\
$^{117}$ University of Kansas, Lawrence, Kansas, United States\\
$^{118}$ University of Liverpool, Liverpool, United Kingdom\\
$^{119}$ University of Science and Technology of China, Hefei, China\\
$^{120}$ University of South-Eastern Norway, Kongsberg, Norway\\
$^{121}$ University of Tennessee, Knoxville, Tennessee, United States\\
$^{122}$ University of the Witwatersrand, Johannesburg, South Africa\\
$^{123}$ University of Tokyo, Tokyo, Japan\\
$^{124}$ University of Tsukuba, Tsukuba, Japan\\
$^{125}$ Universit\"{a}t M\"{u}nster, Institut f\"{u}r Kernphysik, M\"{u}nster, Germany\\
$^{126}$ Universit\'{e} Clermont Auvergne, CNRS/IN2P3, LPC, Clermont-Ferrand, France\\
$^{127}$ Universit\'{e} de Lyon, CNRS/IN2P3, Institut de Physique des 2 Infinis de Lyon, Lyon, France\\
$^{128}$ Universit\'{e} de Strasbourg, CNRS, IPHC UMR 7178, F-67000 Strasbourg, France, Strasbourg, France\\
$^{129}$ Universit\'{e} Paris-Saclay, Centre d'Etudes de Saclay (CEA), IRFU, D\'{e}partment de Physique Nucl\'{e}aire (DPhN), Saclay, France\\
$^{130}$ Universit\'{e}  Paris-Saclay, CNRS/IN2P3, IJCLab, Orsay, France\\
$^{131}$ Universit\`{a} degli Studi di Foggia, Foggia, Italy\\
$^{132}$ Universit\`{a} del Piemonte Orientale, Vercelli, Italy\\
$^{133}$ Universit\`{a} di Brescia, Brescia, Italy\\
$^{134}$ Variable Energy Cyclotron Centre, Homi Bhabha National Institute, Kolkata, India\\
$^{135}$ Warsaw University of Technology, Warsaw, Poland\\
$^{136}$ Wayne State University, Detroit, Michigan, United States\\
$^{137}$ Yale University, New Haven, Connecticut, United States\\
$^{138}$ Yildiz Technical University, Istanbul, Turkey\\
$^{139}$ Yonsei University, Seoul, Republic of Korea\\
$^{140}$ Affiliated with an institute covered by a cooperation agreement with CERN\\
$^{141}$ Affiliated with an international laboratory covered by a cooperation agreement with CERN.\\

\end{flushleft} 